\documentclass[amsmath,superscriptaddress,showpacs,aps,prb,reprint,floatfix]{revtex4-1}
\usepackage{graphicx, color, xfrac}
\usepackage{amsthm, amsmath, amssymb, bm, enumerate}
\usepackage[dvips]{epsfig}
\usepackage[dvips]{epsfig}
\usepackage[normalem]{ulem}

\newcommand{\Nbar}{{\bar{N}}}
\newcommand{\Mbar}{{\bar{M}}}
\newcommand{\ent}{\mathrm{e}}
\newcommand{\zero}{\mathrm{zero}}
\newcommand{\osc}{\mathrm{osc}}
\newcommand{\Tr}{\mathrm{Tr}}
\newcommand{\rt}{{\tilde{r}}}
\newcommand{\pit}{{\tilde{\pi}}}

\begin{document}

\title{Entanglement spectra between coupled Tomonaga-Luttinger liquids: \\Applications to ladder systems and topological phases}
\author{Rex Lundgren}
\affiliation{Department of Physics, The University of Texas at Austin, Austin, TX 78712, USA}
\author{Yohei Fuji}
\affiliation{Institute for Solid State Physics, University of Tokyo, Kashiwa 277-8581, Japan}
\author{Shunsuke Furukawa}
\affiliation{Department of Physics, University of Tokyo, 7-3-1 Hongo, Bunkyo-ku, Tokyo 113-0033, Japan}
\author{Masaki Oshikawa}
\affiliation{Institute for Solid State Physics, University of Tokyo, Kashiwa 277-8581, Japan}

\date{\today}


\begin{abstract}
We study the entanglement spectrum (ES) and entropy between two coupled Tomonaga-Luttinger liquids (TLLs) on parallel periodic chains. 
This problem gives access to the entanglement properties of various interesting
systems, 
such as spin ladders as well as two-dimensional topological phases.
By expanding interchain interactions to quadratic order in bosonic fields, we are able to calculate the ES for both gapped and gapless systems using only methods for free theories. 
In certain gapless phases of coupled non-chiral TLLs, 
we interestingly find an ES with a dispersion relation proportional to the square root of the subsystem momentum, 
which we relate to a long-range interaction in the entanglement Hamiltonian. 
We numerically demonstrate the emergence of this unusual dispersion in a model of hard-core bosons on a ladder. 
In gapped phases of coupled non-chiral TLLs, which are relevant to spin ladders and topological insulators, 
we show that the ES consists of linearly dispersing modes, which resembles the spectrum of a single-chain TLL but is characterized by a modified TLL parameter. 
Based on a calculation for coupled chiral TLLs, 
we are also able to provide a very simple proof for the correspondence between the ES and the edge-state spectrum in quantum Hall systems 
consistent with previous numerical and analytical studies. 
\end{abstract}
\maketitle

\section{Introduction}


Quantum entanglement has been found to be useful in characterizing
topological states of matter,
which is not possible with conventional local order parameters.
For example, 
the topological entanglement entropy (TEE)\cite{PhysRevLett.96.110404,PhysRevLett.96.110405} 
encodes information of topolocically ordered phases.
The TEE has been numerically and analytically calculated in many exotic systems 
including fractional quantum Hall systems\cite{PhysRevLett.98.060401,Zozulya:prb07,Laeuchli:njp10} 
and quantum spin liquids.\cite{PhysRevA.71.022315,PhysRevB.75.214407,PhysRevLett.103.261601,Zhang:prb12,2012NatPh...8..902J, PhysRevLett.109.067201} 
However, TEE does not give the complete picture of the entanglement.
The entanglement spectrum (ES),
as proposed by Li and Haldane,\cite{Li:prl08} 
contains more information than the TEE and is also a powerful tool
in studying topological phases.
A large amount of focus of the investigation of ES has been with a real space partition in various systems. 
Those systems include quantum Hall systems,\cite{Li:prl08, Thomale_AC:prl10, Lauchli:prl10, Sterdyniak:prb12, Dubail:prb12,PhysRevLett.108.256806,PhysRevB.88.155307} 
topological insulators,\cite{PhysRevB.82.241102, PhysRevLett.104.130502,Kargarian:prb10, PhysRevB.84.195103,PhysRevB.87.035119} 
fractional Chern insulators,\cite{PhysRevX.1.021014} 
symmetry-protected topological phases,\cite{Pollmann:prb10, Turner:prb11, Fidkowski:prb11}
quantum spin chains\cite{Calabrese:pra08, Pollmann:njp10, Franchini:2010kq, 2012JSMTE..08..011A,PhysRevLett.108.227201,2013arXiv1303.0741L,2013arXiv1307.7718G}
and ladders,\cite{Poilblanc:prl10,2011EL.....9650006P,Lauchli:prb12,2012JSMTE..11..021S,Tanaka:pra12, Lundgren,Fradkin_Ladder} 
and other spin and fermionic systems.\cite{PhysRevLett.105.080501, Lou:prb11, Cirac:prb11, Santos:prb13, Deng:prb11, 2013NJPh...15e3017S,PhysRevLett.107.157001,2013arXiv1307.1486G}.


For the system's ground state $|\Psi\rangle$, 
the ES is obtained by first partitioning the system into a subregion $A$ and the rest $\bar{A}$ 
and then forming the reduced density matrix $\rho_A=\Tr_{\bar{A}}~|\Psi\rangle\langle\Psi|$ on $A$. 
The reduced density matrix is conventionally written in the thermal form $\rho_A=e^{-H_\ent}$, 
where $H_\ent$ is referred to as the entanglement Hamiltonian. 
The ES is simply the full set of eigenvalues of the entanglement Hamiltonian.

%
\begin{figure}
\begin{center}
\includegraphics[width=0.5\textwidth]{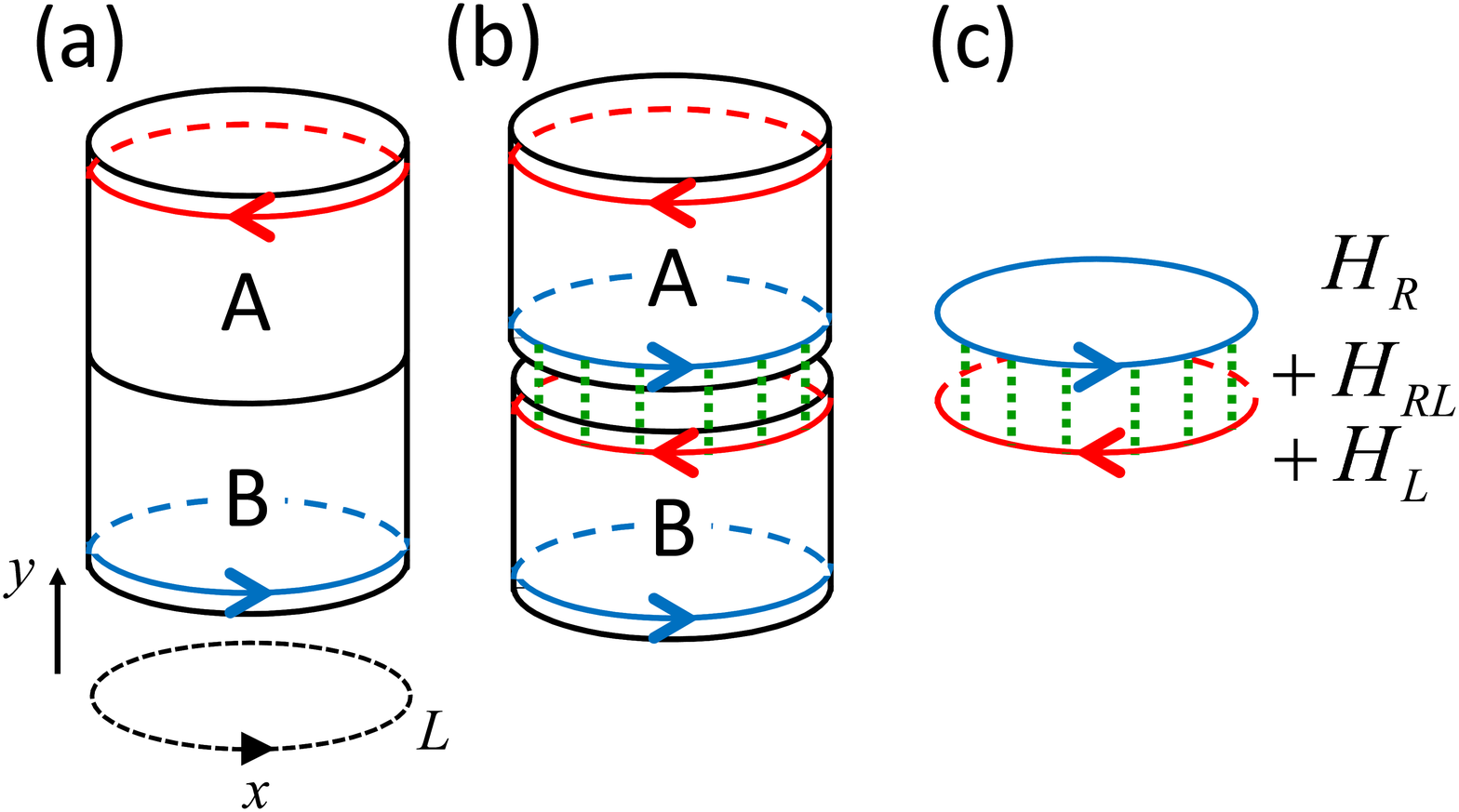}
\end{center}
\caption{(Color online) 
``Cut and glue'' approach\cite{Qi:prl12} applied to a quantum Hall state. 
(a) We define the system on a cylinder of circumference $L$ and 
aim to study the entanglement between two halves of the cylinder, $A$ and $B$. 
(b) To this end,  we physically cut the system into $A$ and $B$, 
obtaining gapless modes described by chiral TLLs, $H_R$ and $H_L$, on the new edges. 
We then glue them along the edges by switching on an interaction that couples $A$ and $B$. 
(c) The entanglement problem of the quantum Hall state can then be reduced to 
the problem of entanglement between the two coupled chiral TLLs. 
}
\label{fig:cylinder}
\end{figure}


A key result found in many topological phases and some other states (often with an energy gap)
has been the remarkable correspondence between the ES and the edge-state spectrum. 
In two-dimensional (2D) topological phases, the ES has been found to exhibit gapless structures 
which are related to the conformal field theory (CFT) describing low-energy edge excitations.\cite{Li:prl08, Thomale_AC:prl10, Lauchli:prl10, Sterdyniak:prb12, Dubail:prb12,
PhysRevB.82.241102,PhysRevLett.104.130502,Kargarian:prb10, PhysRevX.1.021014} 
An analytic proof for this correspondence has been proposed by Qi, Katsura, and Ludwig\cite{Qi:prl12} using the ``cut and glue'' approach. 
For model wave functions of quantum Hall systems, different proofs have been proposed 
by using clustering properties\cite{PhysRevB.84.205136} and a connection to CFT.\cite{Dubail_proof:prb12}
A geometric proof for two- and three-dimensional topological phases has also been put forward.\cite{PhysRevB.86.045117} 
In the approach of Qi {\it et al.},\cite{Qi:prl12}  
it was suggested to start from two pieces $A$ and $B$ of a topological phase which both support gapless chiral edge states, 
and then to glue them along the edges by switching on an interaction Hamiltonian that couples $A$ and $B$ as in Fig.~\ref{fig:cylinder}. 
Restricting the analysis to a coupling between the gapless edge modes only, 
boundary CFT was applied to prove the correspondence 
between the entanglement Hamiltonian and the single chiral edge Hamiltonian. 


The ``cut and glue'' approach suggests that the entanglement properties of topological phases 
are closely related to those of coupled one-dimensional (1D) gapless systems. 
We now give a quick review of results for the ES of coupled 1D systems (``ladders''),  
which have been shown to exhibit various interesting features depending crucially on the interchain couplings. 
Poilblanc\cite{Poilblanc:prl10} numerically calculated the ES in gapped phases of a spin-$\frac12$ Heisenberg ladder, 
where the entanglement cut was introduced between the two periodic chains. 
He observed that the ES remarkably resembles the energy spectrum of a single Heisenberg chain. 
Peschel and Chung\cite{2011EL.....9650006P} and L\"auchli and Schliemann\cite{Lauchli:prb12} 
explained this result using perturbation theory from the strong-rung-coupling limit. 
They also generalized it to the model with an XXZ anisotropy 
and showed that the entanglement Hamiltonian is given by an XXZ chain 
with its anisotropy modified from the physical single-chain Hamiltonian. 
Schliemann and L\"auchli\cite{2012JSMTE..11..021S} also studied the cases of higher-spin ladders (such as spin-$1$), and again found that in the isotropic case, the entanglement Hamiltonian is proportional to the single-chain Hamiltonian. 
We note that unlike the spin-$\frac12$ case, 
the spectrum of the single-chain Hamiltonian is gapped for certain spin lengths. 
ES in a variety of coupled 1D systems, including fermionic and spin ladders, 
was studied by Chen and Fradkin.\cite{Fradkin_Ladder} 
Their result for gapped coupled systems supported the correspondence between the ES and the single-chain spectrum. 
They also investigated a system of two Tomonaga-Luttinger liquids (TLLs) coupled by a marginal interaction, 
and interestingly found an ES with a flat dispersion. 
These results open the question, what condition is necessary for finding the correspondence between the ES and the single-chain spectrum. 
Lundgren, Chua, and Fiete\cite{Lundgren} studied the ES of the Kugel-Khomskii model, 
which can be regarded as a spin ladder system, with spin degrees of freedom on one leg and orbital degrees of freedom on the other. 
In that work an entanglement gap was found at the gapless $SU(4)$ point. 
The origin of this gap and the structure of the levels below the entanglement gap, 
which could have universal features, has yet to be explained.


In this paper, we study the ES between two coupled (chiral or non-chiral) TLLs on parallel periodic chains 
for a variety of interchain interactions. 
In addition to having direct applications to ladder systems, 
this problem is also closely related to the entanglement properties of 2D topological phases via the ``cut and glue'' picture of Fig.~\ref{fig:cylinder}. 
By expanding interchain interactions to quadratic order in bosonic fields, 
we are able to obtain the ES in both gapped and gapless phases using only methods for free theories. 
This is similar to the approach of Chen and Fradkin\cite{Fradkin_Ladder} for a gapless phase  
but extends it to a wider variety of phases of coupled TLLs. 
We also carefully treat zero modes, which turn out to be important to the full structure of the ES. 
Based on the calculation for coupled chiral TLLs, we provide a simple proof 
for the correspondence between the ES and the edge-state spectrum in quantum Hall systems consistent with previous numerical and analytical studies. 
Using this result, we are easily able to obtain the TEE, 
which is equal to $\mathrm{log} \sqrt{q}$, for the quantum Hall state at the filling fraction $\nu=1/q$.\cite{PhysRevLett.96.110404, Fradkin_TEE, 2007JSP...126.1111F}
In gapped phases of coupled non-chiral TLLs, which are relevant to spin ladders and time-reversal-invariant topological insulators,\cite{Kane:prl05} 
we show that the ES consists of linearly dispersing modes, 
which resembles the spectrum of a single-chain TLL but is characterized by a modified TLL parameter. 
When the system becomes partially gapless (either in the symmetric or antisymmetric channels of bosonic fields), 
we interestingly find an ES with a dispersion relation proportional to the square root of the subsystem momentum, 
which we relate to a certain long-range interaction in the entanglement Hamiltonian. 
We numerically demonstrate the emergence of this unusual dispersion in a model of hard-core bosons on a ladder. 


While the edge-state picture has been widely used for ES in {\it gapped} systems, 
few works have successfully addressed the universal features of ES in {\it gapless} systems. 
For 1D critical systems, CFT has been applied to reveal the universal distribution function of the ES.\cite{Calabrese:pra08}
In that case, the partition is a block of some finite length in a longer finite- or infinite-length chain. 
There have been some works for understanding the ES of gapless systems with broken continuous symmetry.\cite{2013PhRvL.110z0403A,2013arXiv1307.6592K} 
Another avenue to explore entanglement in gapless systems is to partition the system in momentum space instead of real space\cite{PhysRevLett.105.116805, PhysRevLett.110.046806}; 
however such work is beyond the scope of this paper. 
Gapless phases of coupled TLLs investigated in Ref.~\onlinecite{Fradkin_Ladder} and here provide novel examples
in which critical correlations manifest themselves in unusual dispersion relations in the ES. 


The rest of the paper is organized as follows. 
In Sec.~\ref{sec:chTLL}, we consider a system of two coupled chiral TLLs 
and show that the ES is proportional to the spectrum of a single chiral TLL.  
This result gives a simple proof for the correspondence between the edge states and the ES in quantum Hall states. 
In Sec.~\ref{sec:nchTLL}, we consider two coupled non-chiral TLLs, 
which are relevant to spin ladders, Hubbard chains, and topological insulators. 
We calculate the ES in both gapped and gapless phases 
and show that it has a variety of low-energy features depending on the phase. 
In Sec.~\ref{sec:numerics}, we numerically demonstrate some of the predictions of Sec.~\ref{sec:nchTLL} 
in a model of hard-core bosons on a ladder. 
Finally, in Sec.~\ref{sec:summary}, we present our summary, conclusions and open questions.

\section{Two coupled chiral Tomonaga-Luttinger liquids} \label{sec:chTLL}

\subsection{Introduction of model}\label{sec.introTLL}


In this section, we study the ES between two coupled chiral TLLs. 
This provides the simplest illustration of our approach. 
Furthermore, this problem is closely related to the ES in quantum Hall states, as we explain below. 

To be specific, we consider the quantum Hall states at filling faction $\nu=\frac{1}{q}$, 
where $q$ is odd for fermions and even for bosons. 
These systems are described by the $(2+1)$-dimensional Chern-Simons gauge theory with the action
\begin{equation}\label{eq:Chern-Simons}
 S=-\int\mathrm{d}^3x \frac{q}{4\pi} a_{\mu}\partial_{\nu}a_{\lambda}\epsilon^{\mu\nu\lambda}, 
\end{equation}
where $x^\mu=(t,x,y)$ is a $(2+1)$-dimensional coordinate and $\epsilon^{\mu\nu\lambda}$ is a Levi-Civita symbol. 
The gauge field $a^\mu$ is related to the particle current $j^\mu$ via $j^\mu=\frac{1}{2\pi} \partial_\nu a_\lambda \epsilon^{\mu\nu\lambda}$. 

We define the system on a cylinder of circumference $L$ 
and study the ES between the upper and lower halves of the system as in Fig.~\ref{fig:cylinder}(a). 
To this end, following the ``cut and glue'' approach of Qi {\it et al.},\cite{Qi:prl12} 
we first physically cut the system as in Fig.~\ref{fig:cylinder}(b). 
Along the new edges of the two subsystems, 
there appear 1D right- and left-moving chiral modes described by the chiral TLL Hamiltonians
\begin{equation}\label{eq:Edge_Ham}
 H_{R/L} = \int_0^L \mathrm{d}x\frac{q v_0}{4\pi} (\partial_x\phi_{R/L})^2,
\end{equation}
where $v_0$ is the velocity. 
These Hamiltonians are derived from the Chern-Simons action \eqref{eq:Chern-Simons} 
by imposing certain gauge-fixing conditions at the edges.\cite{Quantum_Field_Theory} 
Here, the bosonic field $\phi_{R/L}(x)$ is related to the particle density fluctuation $\rho_{R/L}(x)$ (relative to the ground state) 
via $\rho_{R/L}=\frac{1}{2\pi} \partial_x \phi_{R/L} $. 
The annihilation operators of an original fermionic/bosonic particle at the edges are given by
\footnote{For fermionic systems, in fact, some modifications are necessary 
in order to have the anticommutation relations of these operators. 
One simple way to do it is to multiply the factors 
$\exp\left( i\frac{\pi}{2}N_L+i\frac{\pi}{4} \right)$ and $\exp\left( -i\frac{\pi}{2}N_R-i\frac{\pi}{4} \right)$ 
to the expressions of $\psi_R$ and $\psi_L$, respectively. 
In this case, the argument of the cosine term in Eq.~\eqref{eq:Total_Ham} is shifted by $\frac{\pi}{2}(N_L+N_R)$. 
This does not change the subsequent argument by focusing on the sector of fixed total particle number $N_L+N_R=0$.} 
\begin{equation}\label{eq:psi_phi}
 \psi_R = \frac{1}{\sqrt{2\pi}} e^{iq\phi_R}, ~~
 \psi_L = \frac{1}{\sqrt{2\pi}} e^{-iq\phi_L}. 
\end{equation}

We now consider the Hamiltonian $H(\lambda)=H_A+H_B+\lambda H_{AB}$, 
which connects between the decoupled system ($\lambda=0$) and the original system ($\lambda=1$). 
Here, $H_A$ and $H_B$ describe $A$ and $B$ regions, respectively, and 
$H_{AB}$ describes the coupling between them. 
At $\lambda = 0$, the system is decoupled into two
independent quantum Hall systems on regions $A$ and $B$.
Then gapless chiral TLLs appear as the edge states of the
two systems.
When the coupling $\lambda H_{AB}$ is small enough with respect
to the bulk gap, it can be regarded as a perturbation
to the gapless edge modes.
In particular, when $\lambda H_{AB}$ is a relevant perturbation
in the renormalization group (RG) sense, the coupling
$\lambda$ effectively increases as the energy scale is lowered.
If the RG flow simply connects the decoupled limit to the
strong coupling limit, the low-energy limit of the system
with an arbitrary small coupling may be considered
essentially the same as that of the original system with $\lambda=1$.
Thus the entanglement between the two regions $A$ and $B$
in the original system could be
essentially understood in terms of the entanglement
between the two coupled chiral TLLs, as pointed out by
Qi {\it et al.}\cite{Qi:prl12}


The bosonic fields $\phi_{R/L}$ in Eq.~\eqref{eq:Edge_Ham} have the mode expansions
\begin{subequations}\label{eq:Mode_phi}
\begin{align}
 \phi_R =& \phi_{R,0}+ 2\pi N_R \frac{x}{L} \notag\\
 &+ \sum_{k>0} \sqrt{\frac{2\pi}{q L |k|}} \left( a_k e^{ikx} + a_k^\dagger e^{-ikx} \right), \\
 \phi_L =& \phi_{L,0}+ 2\pi N_L \frac{x}{L} \notag\\
 &+ \sum_{k<0} \sqrt{\frac{2\pi}{q L |k|}} \left( a_k e^{ikx} + a_k^\dagger e^{-ikx} \right). 
\end{align}
\end{subequations}
Here, $a_k$ with $k=\frac{2\pi j}{L}$~($j\in \mathbb{Z}\setminus\{0\}$) are bosonic operators describing oscillator modes 
and $N_{R/L}\in\mathbb{Z}$ are the changes in the numbers of particles relative to the ground state. 
The zero modes satisfy the commutation relations 
\begin{equation}\label{eq:phi_N_RL}
 [\phi_{R,0}, N_R]=-\frac{i}{q},~~
 [\phi_{L,0}, N_L]=\frac{i}{q}. 
\end{equation}
By inserting Eq.~\eqref{eq:Mode_phi} into Eq.~\eqref{eq:Edge_Ham}, we obtain the energy spectrum of the chiral TLL Hamiltonians as 
\begin{subequations}\label{eq:Edge_Spec} 
\begin{align}
 H_R =&\frac{\pi q v_0}{L} N_R^2+v_0\sum_{k>0} |k| \left( a_k^\dagger a_k+\frac{1}{2} \right), \label{eq:Edge_Spec_R}\\
 H_L =&\frac{\pi q v_0}{L} N_L^2+v_0\sum_{k<0} |k| \left( a_k^\dagger a_k+\frac{1}{2} \right). 
\end{align}
\end{subequations}


We introduce the interedge coupling consisting of particle tunneling and a density-density interaction: 
\begin{equation}
 H_{RL} = \int_0^L dx \left[ g \left( \psi_L^\dagger \psi_R + \psi_R^\dagger \psi_L \right) + U \rho_L \rho_R \right]. 
\end{equation}
Using Eq.~\eqref{eq:psi_phi} and defining 
\begin{equation}
 \phi=\frac{1}{\sqrt{4\pi}} \left( \phi_L+\phi_R \right), ~~
 \theta=\frac{q}{\sqrt{4\pi}} \left( \phi_L-\phi_R \right), 
\end{equation}
the total Hamiltonian $H\equiv H_L+H_R+H_{RL}$ can be recast into a sine-Gordon Hamiltonian
\begin{equation}\label{eq:Total_Ham}
\begin{split}
 H= \int_0^L \mathrm{d}x  \bigg[ 
 &\frac{v}{2} \left( K(\partial_x\theta)^2+\frac{1}{K}(\partial_x\phi)^2 \right) \\
 &+ \frac{g}{\pi} \cos\left( \sqrt{4\pi} q \phi \right) \bigg]
\end{split}
\end{equation}
with 
\begin{equation}
 v=\sqrt{v_0^2 -\left(\frac{U}{2\pi q}\right)^2},~~
 K=\frac{1}{q} \sqrt{\frac{v_0-\frac{U}{2\pi q}}{v_0+\frac{U}{2\pi q}}}. 
\end{equation}
Here $v$ is the renormalized velocity and $K$ is the TLL parameter. 
The cosine term in Eq.~\eqref{eq:Total_Ham} is relevant in the RG
sense when its scaling dimension $q^2 K$ becomes smaller than $2$. 

In the absence of the density-density interaction ($U=0$), the
inter-edge tunneling is RG relevant only for $q=1$
and RG irrelevant for fractional quantum Hall states with $q > 2$ 
(marginal for $q=2$).
However, for sufficiently large $U>0$, the inter-edge tunneling
can be made RG relevant even for $q \geq 2$.
Even when the inter-edge tunneling is RG irrelevant,
there may be a critical value $g_c>0$ such that the system
flows\cite{Qi:prl12}
to the strong tunneling limit when $g>g_c$.
The present 1D formulation would be still valid
in such a case if $g>g_c$ but $g$ is still
sufficiently small with respect to the bulk gap.

\subsection{Free-field description of the coupled system}


In the rest of Sec.~\ref{sec:chTLL}, we aim to calculate the ES in two coupled chiral TLLs described by Eq.~\eqref{eq:Total_Ham}. 
Let us first consider the situation where
the low-energy limit of the system is given by the strong
inter-edge tunneling limit $g \rightarrow \infty$.
This is the case when the inter-edge tunneling is relevant under RG,
or $g>g_c$ even when the infinitesimal inter-edge tunneling is
irrelevant.

Our approach is based on the simple observation that, in the limit
$g \rightarrow \infty$, $\phi$ is locked into the minimum of the
cosine potential. In order to describe the nontrivial entanglement,
we can expand the cosine term around its minimum as
\begin{equation} \label{eq:expansion}
 \frac{g}{\pi} \cos \left( \sqrt{4\pi} q\phi \right) \approx \mathrm{const.}+ \frac{v m^2}{2K}  \left( \phi-\bar{\phi}_0 \right)^2 +\dots, 
\end{equation}
where $\bar{\phi}_0$ is the locking position. 
The total Hamiltonian $H$ then becomes a Klein-Gordon Hamiltonian with a mass gap $vm$. 
Within this approach, one can also study the case of $g=0$ by setting $m=0$. We note that for small but finite $g$, the expansion in Eq. ~(\ref{eq:expansion}) is not well justified (When $g=0$ exactly, however, the expansion is not necessary and the approach is justified again). This is due to the presence of the higher order terms in the expansion and possible tunneling among the minima of the cosine. Therefore, when the bare value of $g$ is small, this expansion should be done after it grows to a sufficiently large value (comparable to $v$) under RG. This means that our approach is valid when the system size $L$ is sufficiently larger than the correlation length $1/m$. Since $H$ is now quadratic in bosonic fields, 
one can use the methods for free theories to address the entanglement properties of the system 
(as done by Chen and Fradkin\cite{Fradkin_Ladder} for a gapless case). 
We note that after the locking, the winding mode for $\phi$ is suppressed, i.e., $N_L+N_R=0$, 
since $\phi$ can no longer have a large spacial dependence. 

Plugging in the mode expansions \eqref{eq:Mode_phi} into the total Hamiltonian $H$, 
we see that $H$ can be decomposed into zero-mode and oscillator parts: 
\begin{equation}
 H=H^{\mathrm{zero}}+H^{\mathrm{osc}},
\end{equation}
where 
\begin{align}
 H^{\mathrm{zero}}=&\frac{v}{2} \left[ \frac{\pi q^2 K}{L} \left(N_L-N_R\right)^2 + \frac{L m^2}{K} \left( \phi_0-\bar{\phi}_0 \right)^2 \right], \label{eq:Hzero}\\
 H^{\mathrm{osc}}=&\frac{v}{2}\sum_{k\neq0} 
  \left( a_k^{\dagger},a_{-k} \right) 
  \begin{pmatrix} A_k & B_k \\ B_k & A_k \end{pmatrix}  
  \begin{pmatrix} a_k \\ a_{-k}^{\dagger} \end{pmatrix} 
\end{align}
with
\begin{align}
 \phi_0 =& \frac{1}{\sqrt{4\pi}} (\phi_{L,0}+\phi_{R,0}),\\
 A_k=&\frac12 \left( qK+\frac{1}{qK} \right) |k|+\frac{m^2}{2qK |k|},\\
 B_k=&\frac12 \left(-qK+\frac{1}{qK} \right) |k|+\frac{m^2}{2qK |k|}. 
\end{align}


Focusing on $H_{\mathrm{osc}}$, we observe that it can be diagonalized via a Bogoliubov transformation 
\begin{equation}
 \begin{pmatrix} a_k \\ a_{-k}^{\dagger} \end{pmatrix}
 =\begin{pmatrix} \cosh \theta_k &  \sinh \theta_k \\ \sinh \theta_k & \cosh \theta_k\\ \end{pmatrix}  
 \begin{pmatrix} b_k \\ b_{-k}^{\dagger} \end{pmatrix}
\end{equation}
with 
\begin{subequations}
\begin{gather}
 \cosh \left( 2\theta_k \right) = \frac{A_k}{\lambda_k},~~\sinh \left( 2\theta_k \right)=-\frac{B_k}{\lambda_k},\\
 \lambda_k=\sqrt{A_k^2-B_k^2}=\sqrt{k^2+m^2}.
\end{gather}
\end{subequations}
We then obtain 
\begin{equation}
 H^{\mathrm{osc}}=v\sum_{k\neq0} \lambda_k \left( b^{\dagger}_k b_k+\frac{1}{2} \right)
\end{equation}
as expected for the Klein-Gordon model with a mass gap $vm$. 
The ground state $|0\rangle$ of $H^{\mathrm{osc}}$ is specified by the condition that $b_k|0\rangle=0$ for all $k$. 


We note that Qi {\it et al.}\cite{Qi:prl12} used boundary CFT to describe the ground state of Eq.~\eqref{eq:Total_Ham}. 
The advantage of our approach lies in its simplicity and versatility. 
Our approach relies only on the methods for free theories and does not require the knowledge of boundary CFT. 
Furthermore, as we will see, our approach can treat both gapped and gapless systems in a unified manner 
while only the gapped case was discussed in Ref.~\onlinecite{Qi:prl12}.

\subsection{Reduced density matrix}\label{sec:chTLL_RDM}

Now we calculate the reduced density matrix $\rho_A$ for the right-movers  
by tracing out the left-movers (or one leg of the ladder). 
Since the Hamiltonian $H$ is decoupled into the zero-mode and oscillator parts, 
the reduced density matrix is of the form 
$\rho_A=\rho_A^\mathrm{zero}\otimes \rho_A^\mathrm{osc}$. 
Below we first calculate the oscillator part $\rho_A^\osc$ and then the zero-mode part $\rho_A^\zero$. 

\subsubsection{Oscillator part}\label{sec:chTLL_RDM_osc}


The oscillator part $\rho_A^\mathrm{osc}$ can be calculated 
by using Peschel's method\cite{2003JPhA...36L.205P} for free particles. 
We first calculate the two-point correlation functions for the chain with right-moving momentum $k>0$. 
Using the ground state $|0\rangle$ of $H^\osc$, a non-zero correlation function is found as
\begin{align}\label{eq:akdak}
 \langle0|a_k^{\dagger}a_k|0\rangle &= \sinh^2 \theta_k = \frac{\cosh(2\theta_k)-1}2. 
\end{align}
We introduce the ansatz
\begin{equation}
 \rho_A^\mathrm{osc} = \frac1{Z_\mathrm{e}^\mathrm{osc}} e^{-H_\mathrm{e}^\mathrm{osc}},~~
 Z_\ent^\osc = \Tr~ e^{-H_\ent^\osc},
\end{equation}
with 
\begin{equation}\label{eq:He_osc_ch}
 H_\mathrm{e}^\mathrm{osc} = \sum_{k>0} w_k \left(a_k^{\dagger}a_k +\frac12 \right). 
\end{equation}
This gives the Bose distribution 
\begin{equation}\label{eq:akdak_ansatz}
 \mathrm{Tr}\left( a_k^\dagger a_k \rho_A^\mathrm{osc} \right) = \frac1{e^{w_k}-1}. 
\end{equation}
Equating Eq.~\eqref{eq:akdak_ansatz} with Eq.~\eqref{eq:akdak}, we obtain the expression of $w_k$ as
\begin{equation}
 w_k=\ln \frac{\cosh(2\theta_k)+1}{\cosh(2\theta_k)-1}. 
\end{equation}


We now focus on the low-energy part of the entanglement Hamiltonian $H_\ent^\osc$. 
Such part gives a high weight in the reduced density matrix $\rho_A^\osc$. 
To this end, we expand $w_k$ for small $k$. 
For $m>0$ and small $|k|$, we find $\cosh(2\theta_k)\approx \frac{m}{2qK |k|}$ and therefore 
\begin{equation}
 w_k \approx \frac{4qK}{m} |k|. 
\end{equation}
Namely, the entanglement spectrum has a linear dispersion at low energies, 
which resembles the spectrum of the edge Hamiltonian $H_R$. 


For $m=0$, by contrast, the dispersion becomes flat: 
\begin{equation}
 w_k = \ln \frac{qK+(qK)^{-1}+2}{qK+(qK)^{-1}-2}.
\end{equation}
A similar flat dispersion in the entanglement spectrum has been found 
by Chen and Fradkin\cite{Fradkin_Ladder} for two coupled non-chiral TLLs in a gapless phase; 
see Sec.~\ref{sec:nchTLL_gapless} for more details. 

\subsubsection{Zero-mode part}\label{sec:chTLL_RDM_zero}

Next we calculate the zero-mode part $\rho_A^\zero$. 
Defining $\Delta\phi_0=\phi_0-\bar{\phi}_0$ and $\pi_0=\sqrt{\pi} q(N_L-N_R)$, 
we can cast our zero-mode Hamiltonian \eqref{eq:Hzero} as
\begin{equation}\label{eq:Hzero_ho}
 H^{\mathrm{zero}} = \frac{vm}{2} \left[ \frac{K}{Lm} \pi_0^2 +\frac{Lm}{K}(\Delta\phi_0)^2 \right].
\end{equation}
Combining the commutation relations \eqref{eq:phi_N_RL} for the zero modes, 
we find $[\Delta\phi_0,\pi_0]=i$, which allows us to identify $\pi_0=-i\frac{\partial}{\partial\Delta\phi_0}$. 
Therefore, Eq.~\eqref{eq:Hzero_ho} can be identified with the Hamiltonian of a harmonic oscillator 
with discrete momentum $\pi_0$. 
The discreteness becomes irrelevant when $\Delta\phi_0$ is sufficiently localized for large $Lm/K$. 
In this case, the ground state $|G\rangle$ in the $\pi_0$ basis can be simply approximated by a gaussian 
\begin{equation}
 \langle \pi_0|G\rangle \propto e^{-\frac{K}{2Lm} \pi_0^2}. 
\end{equation}
Using $N_L+N_R=0$ (due to the suppression of the winding mode in $\phi$), 
this ground state can be written in a Schmidt decomposed form as
\begin{equation}
 |G\rangle = \sum_\Nbar \frac{1}{\sqrt{z}} e^{-\frac{2\pi q^2K}{Lm} \Nbar^2} |N_R=\Nbar \rangle |N_L=-\Nbar \rangle,
\end{equation}
where $z$ is a normalization constant. 
Tracing over the left-moving degrees of freedom, we obtain the reduced density matrix 
\begin{equation}
 \rho_A^\mathrm{zero} = \sum_{N_R} |N_R \rangle \frac{1}{z} e^{-\frac{4\pi q^2K}{Lm} N_R^2} \langle N_R|
\end{equation}
We then finally arrive at the main result for Sec.~\ref{sec:chTLL_RDM}, 
the low-energy expression of the total entanglement Hamiltonian $H_\ent=H_\ent^\zero+H_\ent^\osc$:
\begin{equation}\label{eq:Hent_ch}
 H_\mathrm{e} =\frac{4qK}{m} \left[ \frac{\pi q}{L} N_R^2 + \sum_{k>0} k \left(a_k^{\dagger}a_k + \frac12\right) \right].
\end{equation}
This is precisely proportional to the physical edge Hamiltonian $H_R$ in Eq. \eqref{eq:Edge_Spec_R}, in both zero and oscillator modes. 
This completes the proof for the correspondence between the entanglement spectrum and the edge-state spectrum. 

\subsection{Topological Entanglement Entropy}

The entanglement entropy $S$ is obtained as the thermal entropy of the entanglement Hamiltonian $H_\ent$ 
at the fictitious temperature $T=1$. 
Kitaev and Preskill\cite{PhysRevLett.96.110404} have calculated $S$  
by assuming a general CFT Hamiltonian for $H_\ent$ and using the modular transformation property of the partition function. 
We here obtain $S$ by explicitly calculating the partition function for the entanglement Hamiltonian \eqref{eq:Hent_ch}. 
This provides a simple illustration of the result of Ref.~\onlinecite{PhysRevLett.96.110404} in free-boson $H_\ent$. 

By performing the $\zeta$-function regularization $\zeta(-1)=-1/12$ for the infinite constant term, 
the entanglement Hamiltonian \eqref{eq:Hent_ch} can be rewritten as
\begin{equation}
 H_\mathrm{e} =v_\mathrm{e} \left[ \frac{\pi q}{L} N_R^2 + \sum_{k>0} k a_k^{\dagger}a_k - \frac{\pi}{12L} \right] 
\end{equation}
with $v_\ent=4qK/m$. 
We consider the partition function $Z_\ent (\beta) = \Tr~e^{-\beta H_\ent}$, 
where $\beta=1/T$ is a fictitious inverse temperature. 
Introducing the modular parameter 
\begin{equation}
 \tau = i\tau_2 = i \frac{\beta v_\mathrm{e}}{L},
\end{equation}
the partition function is calculated as
\begin{equation}
\begin{split}
 Z_\ent (\beta) 
 &= \left( \sum_{N_R=-\infty}^\infty e^{-\pi q\tau_2 N_R^2} \right) 
  e^{\frac{\pi}{24}\tau_2} \prod_{j=1}^\infty \left( \sum_{n_j=0}^\infty e^{-2\pi\tau_2 j n_j}\right)\\
 &= \frac{\theta_3(iq\tau_2)}{\eta(i\tau_2)}, 
\end{split}
\end{equation}
where $\theta_3$ is the third Jacobi theta function and $\eta$ is the Dedekind function. 
We are interested in the asymptotic behavior for large $L$, namely, small $\tau_2$. 
Using the modular transformation properties of $\theta_3$ and $\eta$ (see for example Ref. \onlinecite{CFT}), 
we find
\begin{equation}\label{eq:partition}
 Z_\ent (\beta) = \frac{(q\tau_2)^{-\frac12} \theta_3 (i/q\tau_2)}{ \tau_2^{-\frac12} \eta(i/\tau_2)} \approx \frac{1}{\sqrt{q}},  
\end{equation}
where in the last line, we have used the fact that in the small-$\tau_2$ limit, $\theta_3(i/q\tau_2),\eta(i/\tau_2)\approx 1$. 
The entanglement entropy is then obtained as
\begin{equation}
S=\frac{\partial(T \ln Z_\ent(\beta) )}{\partial T} \bigg|_{T=1} \approx -\ln \sqrt{q}
\end{equation} 
which agrees with the well-known result of Ref.~\onlinecite{PhysRevLett.96.110404}. 
The present calculation without an ultraviolet cutoff has yielded only the universal constant 
and did not produce a boundary-law contribution present in generic states. 
The latter is expected to appear by introducing a high-energy cutoff in the entanglement spectrum. 

In the present calculation, we have implicitly assumed that the system is in the ground state of a fixed topological sector. 
In Abelian quantum Hall states, the constant part of $S$ does not depend on the choice of a topological sector. 
However, by superposing the ground states in different topological sectors, 
we in general obtain a different constant in the entanglement entropy. 
The dependence of $S$ on the choice of a ground state in a variety of closed geometries 
has been analyzed by Dong {\it et al.}\cite{Fradkin_TEE} using the Chern-Simons theory. 
Zhang {\it et al.}\cite{Zhang:prb12}  has analyzed this dependence in further detail for a torus geometry 
and proposed to use it to extract quasiparticle statistics and braiding properties. 
Although interesting, it is beyond the scope of this work to extend to such cases.

\section{Two coupled non-chiral Tomonaga-Luttinger liquids} \label{sec:nchTLL}

\subsection{Introduction of model}


In this section, we study the ES between two coupled non-chiral TLLs. 
This setting is relevant to spin ladders and Hubbard chains.  
Furthermore, it has a close relation with 2D time-reversal-invariant topological insulators 
via a ``cut and glue'' approach in Fig.~\ref{fig:cylinder}.  
The ES in gapped phases of spin-$\frac12$ ladders 
have been studied by Poilblanc\cite{Poilblanc:prl10} for the SU$(2)$-symmetric Heisenberg model 
and in Refs.~\onlinecite{2011EL.....9650006P} and \onlinecite{Lauchli:prb12} for the model with XXZ anisotropy. 
For a system of two non-chiral TLLs coupled by a marginal interaction, 
the entanglement entropy and spectrum have been studied by 
Furukawa and Kim\cite{FurukawaKim:prb11prb13} and Chen and Fradkin,\cite{Fradkin_Ladder} respectively. 
Here we consider more general states of coupled TLLs, 
which include all of the above cases as special examples. 


We assume that the two non-chiral TLLs are equivalent and are described by the gaussian Hamiltonian 
\begin{equation}\label{eq:H1_H2}
 H_\nu = \int \mathrm{d}x \frac{v_0}{2} \left[ K \left(\partial_x\theta_\nu\right)^2 + \frac1K \left(\partial_x\phi_\nu\right)^2 \right],~~
 \nu=1,2. 
\end{equation}
where $v_0$ and $K$ are the velocity and the TLL parameter, respectively, in each chain. 
The dual pair of bosonic fields, $\phi_\nu$ and $\theta_\nu$, satisfy the commutation relation 
\begin{equation}
 [\phi_\nu(x), \theta_{\nu'}(x')] = \frac{i}{2} [1+\mathrm{sgn}(x-x')]\delta_{\nu\nu'}.
\end{equation}
The field $\phi_\nu(x)$ is related to the particle density fluctuation $\rho_\nu(x)$ (relative to the ground state) 
via $\rho_\nu(x)=\frac1{2\pi r} \partial\phi_\nu$ 
while $\theta_\nu(x)/\rt$ represents the phase of a fermionic/bosonic operator. 
Here $r$ and $\rt$ are the compactification radii of $\phi_\nu$ and $\theta_\nu$, respectively.  
One can fix these radii to certain values (while keeping the fixed product $r\rt=1/(2\pi)$) to adjust to different normalizations of bosonization. 
For later applications, we set 
\begin{equation}\label{eq:r_rt}
 r=\frac1{2\sqrt{\pi}},~~\rt=\frac1{\sqrt{\pi}}. 
\end{equation}
In this convention, $K<1$ ($K>1$) corresponds to a repulsive (attractive) intrachain interaction in the case of fermions. 


The bosonic fields have the mode expansions 
\begin{subequations}\label{eq:phi_theta_expand}
\begin{align}
 \phi_\nu (x) &=
  \phi_{\nu,0}+2\pi r N_\nu \frac{x}{L} \notag\\
 &+ \sum_{k\ne 0} \sqrt{\frac{K}{2 L|k|}} \left(a_{k\nu}e^{ikx}+ a^{\dagger}_{k\nu} e^{-ikx}\right) \label{eq:mode_ex_non_ch_1} \\
 \theta_\nu (x) &=
  \theta_{\nu,0}+2\pi \tilde{r} M_\nu \frac{x}{L} \notag\\
 &+ \sum_{k\ne 0} \frac{\mathrm{sgn}(-k)}{\sqrt{2KL |k|}}  \left(a_{k\nu}e^{ikx}+a^{\dagger}_{k\nu}e^{-ikx}\right) \label{eq:mode_ex_non_ch_2}.
\end{align}
\end{subequations}
Here, $a_{k\nu}$ with $k=2\pi j/L$ ($j\in \mathbb{Z}\setminus \{0\}$) are bosonic operators describing oscillator modes. 
For bosonic systems, the winding numbers $N_\nu$ and $M_\nu$ are both integers. 
For fermionic systems, they obey the following condition,\cite{Haldane:prl81} called the twisted structure: 
$N_\nu\in\mathbb{Z}$, $N_\nu+2M_\nu\in 2\mathbb{Z}$. 
The winding number $N_\nu$ gives the change in the number of particles in the $\nu$th chain relative to the ground state. 
The constant terms, $\phi_{\nu,0}$ and $\theta_{\nu,0}$, and the winding numbers satisfy the commutation relations 
\begin{equation}\label{eq:phi_theta_M_N}
 \left[ \phi_{\nu,0}, 2\pi \tilde{r} M_{\nu'} \right] = i \delta_{\nu\nu'},~~
 \left[ \theta_{\nu,0}, 2\pi r N_{\nu'} \right] = i \delta_{\nu\nu'}. 
\end{equation}
By inserting expansions \eqref{eq:phi_theta_expand} into Eq.~\eqref{eq:H1_H2}, 
the single-chain Hamiltonians are diagonalized as
\begin{equation}
\begin{split}
 H_\nu = 
 &\frac{2\pi^2 v_0}{L} \left( \tilde{r}^2 K M_\nu^2 + \frac{r^2}K N_\nu^2 \right)\\
 &+v_0\sum_{k\ne 0} |k|\left(a_{k\nu}^\dagger a_{k\nu} +\frac12 \right) .
\end{split}
\end{equation}


For the interchain coupling, we consider the following interactions: 
\begin{equation}
\begin{split}
 H_{12} = 
 \int\mathrm{d}x \bigg[ 
 &g_+ \cos \left(\frac{\phi_1+\phi_2}{r}\right) + g_- \cos \left(\frac{\theta_1-\theta_2}{\rt}\right)\\
 &+ \frac{U}{4\pi^2r^2} \partial_x\phi_1 \partial_x\phi_2 
 \bigg]. 
\end{split}
\end{equation}
Here, the $g_-$ and $U$ terms arise from particle tunneling and a density-density interaction, respectively. 
In the Hubbard chain, the $g_+$ term appears at half-filling and induces a Mott gap.\cite{Giamarchi:book,Gogolin:book} 
In spin-$\frac12$ ladders in the absence of a magnetic field, 
combination of the $g_+$ and $g_-$ terms induces rung-singlet and Haldane phases.\cite{Giamarchi:book,Gogolin:book}


To describe the coupled Hamiltonian $H=H_1+H_2+H_{12}$, 
it is useful to introduce the symmetric and anti-symmetric combinations of bosonic fields as
\begin{equation}
\phi_\pm = \frac{1}{\sqrt{2}} (\phi_1\pm \phi_2),~~
\theta_\pm = \frac{1}{\sqrt{2}} (\theta_1\pm \theta_2).  
\end{equation}
The total Hamiltonian can then be decoupled into sine-Gordon Hamiltonians defined for the symmetric and antisymmetric channels: 
\begin{equation} \label{eq:Hp+Hm}
 H=H_++H_-
\end{equation}
with 
\begin{subequations}\label{eq:sine-Gordon_pm}
\begin{align}
 H_+ = \int \mathrm{d}x \bigg\{ &
 \frac{v_+}{2} \left[ K_+ \left( \partial_x\theta_+ \right)^2 + \frac1{K_+} \left( \partial_x\phi_+ \right)^2 \right] \notag\\
 &+ g_+ \cos \left( \sqrt{2}\phi_+/r \right) \bigg\},\\
 H_- = \int \mathrm{d}x \bigg\{ &
 \frac{v_-}{2} \left[ K_- \left( \partial_x\theta_- \right)^2 + \frac1{K_-} \left( \partial_x\phi_- \right)^2 \right] \notag\\
 &+ g_- \cos \left( \sqrt{2}\theta_-/\rt \right) \bigg\}. 
\end{align}
\end{subequations}
Here the renormalized velocities $v_\pm$ and the TLL parameters $K_\pm$ are given by
\begin{equation}
 v_\pm = v_0 \left( 1\pm \frac{KU}{4\pi^2v_0 r^2} \right)^{\frac12},~~
 K_\pm = K \left( 1\pm \frac{KU}{4\pi^2v_0 r^2} \right)^{-\frac12}. 
\end{equation}
The $g_+$ and $g_-$ terms are relevant in the RG sense 
when their scaling dimensions, $K_+/(2\pi r^2)$ and $1/(2\pi\rt^2 K_-)$ respectively, become smaller than $2$.  

\subsection{Free-field description of the coupled system}

When the cosine terms in Eq.~\eqref{eq:sine-Gordon_pm} become relevant, 
the fields $\phi_+$ and $\theta_-$ are locked into the minima of these terms. 
As done before in Sec.~\ref{sec:chTLL}, we expand these terms to quadratic order around their minima:  
\begin{subequations}
\begin{align}
 &g_+ \cos \left( \sqrt{2}\phi_+/r \right) \notag\\
 &\approx \mathrm{const.}  
 + \frac{v_+ m_+^2}{2K_+} \left( \phi_+ -\bar{\phi}_{+,0} \right)^2 + \dots,\\
 &g_- \cos \left( \sqrt{2}\theta_-/\rt \right)  \notag\\
 &\approx \mathrm{const.} 
 + \frac{v_- K_- m_-^2}{2} \left( \theta_- -\bar{\theta}_{-,0} \right)^2 + \dots,
\end{align}
\end{subequations}
where $\bar{\phi}_{+,0}$ and $\bar{\theta}_{-,0}$ are the locking positions. 
The Hamiltonian $H_\pm$ then becomes
a Klein-Gordon Hamiltonian with a mass gap $vm_\pm$.  
As we have discussed in Sec.~\ref{sec.introTLL},
the above description actually applies even if the (infinitesimal)
cosine terms are irrelevant in the RG sense, provided that
the strong coupling limit $g_\pm \rightarrow \infty$ is
realized.

On the other hand, when either or both
of the cosine terms is RG irrelevant, or if it is fine-tuned
to zero, the system remains gapless.
When one of the modes $\phi_+$ and $\theta_-$ is gapless,
the low-energy limit of the system is a single-component TLL.
Likewise, when the two modes remain gapless, it is  
a two-component TLL.
Such cases can be also treated in the present formulation,
simply by setting $m_+ = 0$ and/or $m_- = 0$.
We emphasize that, the mode that remains gapless still
can contribute to nontrivial entanglement, owing to
the density-density interaction $U$ which mixes the
original variables $\phi_{1,2}$.


Using Eq.~\eqref{eq:phi_theta_expand}, 
we obtain the mode expansions of $\phi_\pm$ and $\theta_\pm$ as
\begin{subequations}\label{eq:phipm_thetapm_expand}
\begin{align}
 \phi_\pm (x) &= \phi_{\pm,0} + \pit_{\pm,0}\frac{x}{L} \notag \\
 &+ \sum_{k\ne 0} \sqrt{\frac{K}{2L|k|}} \left( a_{k,\pm}e^{ikx}+a_{k,\pm}^\dagger e^{-ikx} \right),\\
 \theta_\pm (x) &= \theta_{\pm,0} + \pi_{\pm,0}\frac{x}{L} \notag \\
 &+ \sum_{k\ne 0} \frac{\mathrm{sgn}(-k)}{ \sqrt{2KL |k|} }  \left( a_{k,\pm}e^{ikx}+a_{k,\pm}^\dagger e^{-ikx} \right),
\end{align}
\end{subequations}
where we have defined 
\begin{subequations}
\begin{gather}
 \phi_{\pm,0}=\frac1{\sqrt{2}} \left( \phi_{1,0}\pm\phi_{2,0} \right),~
 \theta_{\pm,0}=\frac1{\sqrt{2}} \left( \theta_{1,0}\pm\theta_{2,0} \right), \\
  \pit_{\pm,0} = \frac{2\pi r}{\sqrt{2}} \left( N_1 \pm N_2 \right),~
 \pi_{\pm,0} = \frac{2\pi\rt}{\sqrt{2}} \left(M_1\pm M_2\right), \\
  a_{k,\pm}=\frac1{\sqrt{2}} \left( a_{k,1} \pm a_{k,2} \right). 
\end{gather}
\end{subequations}
Due to the locking of $\phi_+$ and $\theta_-$, the winding modes of these two fields must be suppressed, 
namely, $\pit_{+,0}=\pi_{-,0}=0$. 
This yields the constraints $N_1=-N_2$ and $M_1=M_2$. 


By inserting the expansions \eqref{eq:phipm_thetapm_expand} (with the condition $\pit_{+,0}=\pi_{-,0}=0$) into the Klein-Gordon Hamiltonians, 
we find that these Hamiltonians can be decoupled into the zero-mode and oscillator parts: 
\begin{equation}
 H_\pm=H_\pm^\zero + H_\pm^\osc.
\end{equation} 
Here the zero-mode part $H_\pm^\zero$ is given by 
\begin{subequations}
\begin{align}
 H_+^\zero &= \frac{v_+}2 \left[ \frac{K_+}{L} \pi_+^2 + \frac{Lm_+^2}{K_+} \left(\Delta\phi_{+,0}\right)^2 \right],\\
 H_-^\zero &= \frac{v_-}2 \left[ \frac1{K_- L} \pit_-^2 + K_- Lm_-^2 \left(\Delta\theta_{-,0}\right)^2 \right],
\end{align}
\end{subequations}
where we have defined $\Delta\phi_{+,0}=\phi_{+,0}-\bar{\phi}_{+,0}$ and 
$\Delta\theta_{-,0}=\theta_{-,0}-\bar{\theta}_{-,0}$. 
Using Eq.~\eqref{eq:phi_theta_M_N}, we obtain the commutation relations
\begin{equation}
 \left[ \Delta\phi_{+,0}, \pi_{+,0} \right] =i,~~
 \left[ \Delta\theta_{-,0}, \pit_{-,0} \right] =i.
\end{equation}
The oscillator part $H_\pm^\osc$ is given by
\begin{equation}\label{eq:Hpmosc}
 H_\pm^\mathrm{osc}=\frac{v_\pm}{2} \sum_{k\neq0} 
  \left( a_{k,\pm}^{\dagger},a_{-k,\pm} \right) 
  \begin{pmatrix} A_{k,\pm} & B_{k,\pm} \\ B_{k,\pm} & A_{k,\pm} \end{pmatrix}  
  \begin{pmatrix} a_{k,\pm} \\ a_{-k,\pm}^\dagger \end{pmatrix} 
\end{equation}
with 
\begin{equation}
\begin{split}
 A_{k,+} &= \frac12 \left( \frac{K_+}{K} + \frac{K}{K_+} \right) |k|+ \frac{Km_+^2}{2K_+|k|}, \\
 B_{k,+} &= \frac12 \left( -\frac{K_+}{K} + \frac{K}{K_+} \right) |k| + \frac{Km_+^2}{2K_+|k|}, \\
 A_{k,-} &= \frac12 \left( \frac{K_-}{K} + \frac{K}{K_-} \right) |k| + \frac{K_-m_-^2}{2K|k|}, \\
 B_{k,-} &= \frac12 \left( -\frac{K_-}{K} + \frac{K}{K_-} \right) |k| - \frac{K_-m_-^2}{2K|k|}. 
\end{split}
\end{equation}
By performing a Bogoliubov transformation
\begin{equation}
 \begin{pmatrix} a_{k,\pm} \\ a_{-k,\pm}^{\dagger} \end{pmatrix}
 =\begin{pmatrix} \cosh \theta_{k,\pm} &  \sinh \theta_{k,\pm} \\ \sinh \theta_{k,\pm} & \cosh \theta_{k,\pm} \\ \end{pmatrix}  
 \begin{pmatrix} b_{k,\pm} \\ b_{-k,\pm}^{\dagger} \end{pmatrix}
\end{equation}
with 
\begin{subequations}
\begin{gather}
 \cosh \left( 2\theta_{k,\pm} \right) = \frac{A_{k,\pm}}{\lambda_{k,\pm}},~~\sinh \left( 2\theta_{k,\pm} \right)=-\frac{B_{k,\pm}}{\lambda_{k,\pm}},\\
 \lambda_{k,\pm}=\sqrt{A_{k,\pm}^2-B_{k,\pm}^2}=\sqrt{k^2+m_\pm^2},  
\end{gather}
\end{subequations}
the oscillator part $H_\pm^\osc$ is diagonalized as
\begin{equation}
 H_\pm^\osc = v_\pm \sum_{k\ne 0} \lambda_{k,\pm} \left( b_{k,\pm}^\dagger b_{k,\pm} + \frac12 \right). 
\end{equation}
The ground state $|0\rangle$ of $H_+^\osc + H_-^\osc$ is specified by the condition that 
$b_{k,\pm}|0\rangle=0$ for all $k\ne 0$. 

\subsection{Reduced density matrix}


Using the ground state of $H=H_+ + H_-$, 
we calculate the reduced density matrix $\rho_A$ for the first chain 
by tracing out the degrees of freedom in the second chain. 
The reduced density matrix can be factorized into the zero-mode and oscillator parts: 
$\rho_A=\rho_A^\zero \otimes \rho_A^\osc$. 


We first consider the zero-mode part $\rho_A^\zero$. 
Using the analogy with a harmonic oscillator as in Sec.~\ref{sec:chTLL_RDM_zero}, 
the ground state $|G\rangle$ of $H_+^\zero+H_-^\zero$ is calculated as
\begin{equation}
\begin{split}
 |G\rangle = &\sum_{(\Mbar,\Nbar)} 
 \frac1{\sqrt{z}} \exp \left[ \frac{4\pi^2 }{L} \left( \frac{\tilde{r}^2 K_+}{m_+} \Mbar^2 + \frac{r^2}{K_-m_-} \Nbar^2 \right) \right]\\
 &\times |(M_1,N_1)=(\Mbar,\Nbar) \rangle |(M_2,N_2)=(\Mbar,-\Nbar)\rangle, 
\end{split}
\end{equation}
where $z$ is a normalization factor. 
From this, the zero-mode part of the entanglement Hamiltonian is calculated as
\begin{equation}\label{eq:Hent_zero_nch}
 H_\ent^\zero = \frac{8\pi^2 }{L} \left( \frac{\tilde{r}^2K_+}{m_+} M_1^2 + \frac{r^2}{K_-m_-} N_1^2 \right). 
\end{equation}


We next calculate the oscillator part $\rho_\ent^\osc$ using Peschel's method\cite{2003JPhA...36L.205P} as in Sec.~\ref{sec:chTLL_RDM_osc}. 
Using the ground state $|0\rangle$ of $H_+^\osc + H_-^\osc$, 
non-zero two-point correlation functions are found as follows: 
\begin{subequations}\label{eq:corr_aa}
\begin{align}
&\langle 0|a_{k,1}^{\dagger}a_{k,1}|0\rangle\notag\\
 &=\frac{1}{2} \left( \sinh^2 \theta_{k,+} + \sinh^2 \theta_{k,-} \right) \notag\\
 &=\frac{1}{4} \left[ \cosh (2\theta_{k,+}) + \cosh (2\theta_{k,-}) \right] -\frac12, \\ 
&\langle 0|a_{k,1}a_{-k,1}|0\rangle \notag\\
 &=\frac{1}{2} \left[ \sinh\theta_{k,+} \cosh\theta_{k,+}+ \sinh\theta_{k,-} \cosh\theta_{k,-} \right] \notag\\ 
 &=\frac{1}{4} \left[ \sinh(2\theta_{k,+})+ \sinh(2\theta_{k,-}) \right].
\end{align}
\end{subequations}
In contrast to Sec.~\ref{sec:chTLL_RDM_osc}, 
``anomalous'' correlation functions, $\langle 0|a_{k,1}a_{-k,1}|0\rangle$, can be non-zero in the present case. 
Taking account of the presence of these correlations, we introduce the ansatz
\begin{equation}
 \rho_A^\osc = \frac1{Z_\ent^\osc} e^{-H_\ent^\osc}, ~~Z_\ent^\osc = \Tr~e^{-H_\ent^\osc}
\end{equation}
with 
\begin{equation}\label{eq:H_ent_ukvk_ak}
 H_\ent^\osc = \frac12 \sum_{k\neq0} 
  \left( a_{k,1}^{\dagger},a_{-k,1} \right) 
  \begin{pmatrix} u_k & v_k \\ v_k & u_k \end{pmatrix}  
  \begin{pmatrix} a_{k,1} \\ a_{-k,1}^\dagger \end{pmatrix} . 
\end{equation}
By performing a Bogoliubov transformation 
\begin{equation}
 \begin{pmatrix} a_{k,1} \\ a_{-k,1}^{\dagger} \end{pmatrix}
 =\begin{pmatrix} \cosh \phi_k &  \sinh \phi_k \\ \sinh \phi_k & \cosh \phi_k \\ \end{pmatrix}  
 \begin{pmatrix} b_{k,1} \\ b_{-k,1}^{\dagger} \end{pmatrix}
\end{equation}
with
\begin{subequations}
\begin{gather}
 \cosh(2\phi_k)=\frac{u_k}{w_k},~~ 
 \sinh(2\phi_k)=-\frac{v_k}{w_k},\\
 w_k=\sqrt{u_k^2-v_k^2}, 
\end{gather}
\end{subequations}
Eq.~\eqref{eq:H_ent_ukvk_ak} is diagonalized as
\begin{equation}\label{eq:H_ent_wk_bk}
 H_\ent^\osc = \sum_{k\ne 0} w_k \left( b_{k,1}^\dagger b_{k,1}+\frac12 \right). 
\end{equation}
Using the Bose distribution 
\begin{equation}
 \Tr~( b_{k,1}^\dagger b_{k,1} \rho_\ent^\osc) = \frac1{e^{w_k}-1}, 
\end{equation}
we obtain the two-point correlation functions as
\begin{subequations}\label{eq:corr_aa_ansatz}
\begin{align}
 \Tr~ ( a_{k,1}^\dagger a_{k,1} \rho_A^\osc ) &= \frac{f_k u_k}{2w_k} - \frac12,\\
 \Tr~ (  a_{k,1} a_{-k,1} \rho_A^\osc )  &= -\frac{f_k v_k}{2w_k}, 
\end{align}
\end{subequations}
where
\begin{equation}
 f_k = \frac{e^{w_k}+1}{e^{w_k}-1} . 
\end{equation}
Equating Eq.~\eqref{eq:corr_aa_ansatz} with Eq.~\eqref{eq:corr_aa}, 
we obtain the expressions for $u_k$ and $v_k$ as
\begin{subequations}
\begin{align}
u_k &= \frac{w_k}{2f_k} \left[ \cosh(2\theta_{k,+})+\cosh(2\theta_{k,-}) \right],\\
v_k &= - \frac{w_k}{2f_k} \left[\sinh(2\theta_{k,+})+\sinh(2\theta_{k,-})\right], 
\end{align}
\end{subequations}
where 
\begin{subequations}
\begin{align}
 f_k 
 &= \frac12 \big\{ \left[\cosh(2\theta_{k,+})+\cosh(2\theta_{k,-})\right]^2 \notag\\
 &~~~~~~- \left[\sinh(2\theta_{k,+})+\sinh(2\theta_{k,-})\right]^2  \big\}^{\frac12},\\
 w_k &= \ln \frac{f_k+1}{f_k-1}. 
\end{align}
\end{subequations}

Here, $u_k$, $v_k$, and $w_k$ are obtained as complicated functions of $k$. 
Below we expand these functions for small $|k|$ and discuss the low-energy expression of $H_\ent^\osc$ in four different cases.

\subsubsection{Case of $m_\pm >0$} \label{sec:nchTLL_cs_gapped}

We first consider the case in which both the symmetric and antisymmetric channels are gapped. 
For small $|k|$, we find 
\begin{equation}
\begin{split}
 &\cosh(2\theta_{k,+}) \approx -\sinh(2\theta_{k,+}) \approx \frac{Km_+}{2K_+|k|},\\
 &\cosh(2\theta_{k,-}) \approx \sinh(2\theta_{k,-}) \approx \frac{K_-m_-}{2K|k|}. 
\end{split}
\end{equation}
From these, we obtain the small-$|k|$ expressions of $u_k$, $v_k$, and $w_k$ as 
\begin{subequations}\label{eq:ukvkwk_gapped}
\begin{align}
 u_k &\approx \frac{v_\ent|k|}{2} \left(\frac{K_\ent}{K} + \frac{K}{K_\ent}\right),\\
 v_k &\approx \frac{v_\ent|k|}{2} \left(-\frac{K_\ent}{K} + \frac{K}{K_\ent}\right),\\
 w_k &\approx v_\ent |k|, \label{eq:wk_linear}
\end{align}
\end{subequations}
where we have defined
\begin{equation}\label{eq:ve_Ke}
 v_\ent = 4 \sqrt{\frac{K_+}{K_- m_+m_-}},~~
 K_\ent = \sqrt{\frac{K_+K_-m_-}{m_+}}. 
\end{equation}
We have obtained a low-energy linear dispersion $w_k$ with a velocity $v_\ent$ for the bosonic modes in Eq.~\eqref{eq:H_ent_wk_bk}, 
which resembles the spectrum of a single TLL. 
Using Eq.~\eqref{eq:ve_Ke}, 
the zero-mode part in Eq.~\eqref{eq:Hent_zero_nch} can be rewritten as
\begin{equation}\label{eq:Hent_zero_nch_gapped}
 H_\ent^\zero = \frac{2\pi^2 v_\ent}{L} \left( \tilde{r}^2 K_\ent M_1^2 + \frac{r^2}{K_\ent} N_1^2 \right).
\end{equation}
Combining the zero-mode and oscillator parts, we find that 
the total entanglement Hamiltonian $H_\ent=H_\ent^\zero+H_\ent^\osc$ can be recast at low energies into 
a TLL Hamiltonian having a renormalized TLL parameter $K_\ent$: 
\begin{equation}
 H_\ent = \int \mathrm{d}x \frac{v_\ent}{2} \left[ K_\ent (\partial_x \theta_1)^2 + \frac1{K_\ent} (\partial_x \phi_1)^2 \right]  . 
\end{equation}
One can easily confirm that the insertion of the mode expansions \eqref{eq:phi_theta_expand} to this equation
precisely reproduces the zero-mode part \eqref{eq:Hent_zero_nch_gapped} 
and the oscillator part \eqref{eq:H_ent_ukvk_ak} with $u_k$ and $v_k$ given by Eq.~\eqref{eq:ukvkwk_gapped}. 

For the spin-$\frac12$ Heisenberg ladder studied by Poilblanc,\cite{Poilblanc:prl10} 
both the renormalized TLL parameter $K_\ent$ and the original single-chain one $K$ is fixed to $1/2$ [in the normalization of Eq.~\eqref{eq:r_rt}]
because of the SU$(2)$ symmetry present in this system.  
In this case, our calculation indicates that $H_\ent$ is directly proportional to $H_1$ at low energies, 
in consistency with the numerical result of Poilblanc.\cite{Poilblanc:prl10}
Without such a special symmetry, however, $K_\ent$ is in general different from $K$. 
This indicates that the correspondence between $H_\ent$ and $H_1$ is slightly violated in the value of the TLL parameter. 
This explains the result of Refs.~\onlinecite{2011EL.....9650006P} and \onlinecite{Lauchli:prb12} 
in a spin-$\frac12$ XXZ ladder, 
in which the entanglement Hamlitonian had renormalized XXZ anisotropy that differed from the single-chain Hamiltonian, 
from a field-theortical point of view.

The boundary CFT approach of Qi {\it et al.}\cite{Qi:prl12} can also be used to explain Poilblanc's result 
by viewing the ladder system as two copies of coupled chiral TLLs.\footnote{We thank X.-L. Qi for sharing this information with us.}  
Namely, the right (left) mover of the first chain is coupled with the left (right) mover of the second chain. 
Such a picture, however, is applicable only in the condition of $U=0$ and $m_+=m_-$. 
Our calculation has revealed that the violation of this condition leads to $K_\ent \ne K$. 

Using the ``cut and glue'' approach in Fig.~\ref{fig:cylinder}, 
the present calculation can also be used to discuss the ES in 2D time-reversal-invariant topological insulators.\cite{Kane:prl05} 
In the presence of interactions, the edge modes of these systems are described by a helical TLL \cite{Wu:prl06, Xu:prb06}; 
the coupling between the two edges is then described as in Ref.~\onlinecite{Tada:prb12}. 
In non-interacting systems, the entanglement Hamiltonian is also described by non-interacting fermions; 
in this case, both $K$ and $K_\ent$ must be fixed to $1$ [in the normalization of Eq.~\eqref{eq:r_rt}]. 
With interactions, $K$ deviates from $1$, as seen in a recent numerical simulation of the Kane-Mele-Hubbard model.\cite{Hohenadler:prb12}
In this interacting case, our result indicates $K_\ent \ne K$.

\subsubsection{Case of $m_+ =0,~m_->0$} \label{sec:nchTLL_c_gapped}

We next consider the case in which the symmetric channel is gapless and the antisymmetric channel is gapped. 
For small $|k|$, we find 
\begin{equation}
\begin{split}
 &\cosh(2\theta_{k,+}) = \frac12 \left( \frac{K_+}{K} + \frac{K}{K_+} \right), \\
 &\sinh(2\theta_{k,+}) = \frac12 \left( \frac{K_+}{K} - \frac{K}{K_+} \right), \\
 &\cosh(2\theta_{k,-}) \approx \sinh(2\theta_{k,-}) \approx \frac{K_-m_-}{2K|k|},
\end{split}
\end{equation}
from which we obtain the small-$|k|$ expressions of $u_k$, $v_k$, and $w_k$ as 
\begin{subequations}\label{eq:ukvkwk_c_gapped} 
\begin{align} 
 u_k &\approx \frac{v_\ent |k|}{2} \left( \frac{K_+}{K} + \frac{K}{K_+} \right) + \frac{2K_+}{K},\\
 v_k &\approx \frac{v_\ent |k|}{2} \left( -\frac{K_+}{K} + \frac{K}{K_+} \right) - \frac{2K_+}{K},\\
 w_k &\approx 2\sqrt{v_\ent |k|}, \label{eq:wk_c_gapped}
\end{align}
\end{subequations}
with $v_\ent=\frac{4K_+}{K_-m_-}$. 
Remarkably, we have obtained a dispersion relation $w_k$ proportional to $\sqrt{k}$. While we are unaware of a quantum many-body system with such an energy spectrum, we note that a square root dispersion appears in deep water waves. \cite{DEEP} This sharply contrasts with the linear dispersions obtained in Sec.~\ref{sec:chTLL} and Sec.~\ref{sec:nchTLL_cs_gapped}. 
Such a non-analytic dispersion cannot be described by
any \emph{local} entanglement Hamiltonian.
Indeed, the entanglement Hamiltonian in this case is shown to
include a \emph{non-local} interaction, as follows.

By discarding the first term in Eq.~\eqref{eq:ve_Ke} (because $M_{1,2}=0$ in the ground state of the coupled system), 
we obtain the zero-mode part $H_\ent^\zero$ as
\begin{equation}\label{eq:Hent_zero_nch_c_gapped} 
 H_\ent^\zero = \frac{8\pi^2 r^2}{K_-Lm_-}  N_1^2 = \frac{2\pi^2 v_\ent r^2}{K_+L} N_1^2. 
\end{equation}
Combining the zero-mode and oscillator parts, 
the total entanglement Hamiltonian $H_\ent=H_\ent^\zero+H_\ent^\osc$ can be recast into
\begin{equation}\label{eq:Hent_c_gapped}
\begin{split}
 &H_\ent = \int \mathrm{d}x 
  \frac{v_\ent}{2} \left[ K_+(\partial_x\theta_1)^2 + \frac1{K_+} (\partial_x\phi_1)^2 \right] \\
  &- \frac{2K_+}{\pi} \int\!\! \int dx dx' \partial_x\theta_1(x) \partial_{x'}\theta_1(x') \ln \big| e^{i\frac{2\pi}{L}x}-e^{i\frac{2\pi}{L}x'} \big| .
\end{split}
\end{equation}
Here, to obtain the second line, we used the identity
\begin{equation}
\begin{split}
 &\lim_{\alpha\searrow 0} \frac1L \sum_{k\ne 0} |k| e^{ik(x-x')-\alpha|k|} \\
 &= -\frac1\pi \partial_x\partial_{x'} \ln \big| e^{i\frac{2\pi}{L}x}-e^{i\frac{2\pi}{L}x'} \big|, 
\end{split}
\end{equation}
where $\alpha$ is a short-distance cutoff for regularization. 
One can check that the insertion of the mode expansions \eqref{eq:phi_theta_expand} with $M_1=0$ into Eq.~\eqref{eq:Hent_c_gapped}
precisely reproduces the zero-mode part \eqref{eq:Hent_zero_nch_c_gapped} 
and the oscillator part \eqref{eq:H_ent_ukvk_ak} with $u_k$ and $v_k$ given by Eq.~\eqref{eq:ukvkwk_c_gapped}. 
The first line of Eq.~\eqref{eq:Hent_c_gapped} has the form of a TLL Hamiltonian, 
which resembles $H_1$ but has a renormalized TLL parameter $K_+$. 
A new, very unusual feature is found in the second line of Eq.~\eqref{eq:Hent_c_gapped}, 
where the field $\partial_x\theta_1$ has a long-range interaction proportional to the logarithm of the chord distance on a unit circle. 
A similar logarithmic potential is also found in the field-theoretical representation of the TLL ground state wave function.\cite{Fradkin1993667,Stephan:prb09, FurukawaKim:prb11prb13}
We note that in the single-chain TLL theory with the Hamiltonian $H_1$ in Eq.~\eqref{eq:H1_H2}, 
the field $\partial_x\theta_1$ is related to the local current $j_1(x)$ via $j_1=-v_0\partial_x\theta_1$. We also note that such a remarkable dispersion seen in Eq.~(\ref{eq:wk_c_gapped}) cannot be obtained by simple power law terms. \cite{2006JPhA...39.2161I}

\subsubsection{Case of $m_+ >0,~m_-=0$} \label{sec:nchTLL_s_gapped}

Similarly to the previous case, the dispersion $w_k$ in the case of $m_+ >0,~m_-=0$ 
is obtained as
\begin{equation}
 w_k \approx 2 \sqrt{v_\ent |k|}, ~~v_\ent=\frac{4K_+}{K_-m_+}. 
\end{equation}
The entanglement Hamiltonian is given by
\begin{equation}\label{eq:Hent_c_gapped_2}
\begin{split}
 &H_\ent = \int \mathrm{d}x 
  \frac{v_\ent}{2} \left[ K_-(\partial_x\theta_1)^2 + \frac1{K_-} (\partial_x\phi_1)^2 \right] \\
  &- \frac{2}{\pi K_-} \int\!\! \int dx dx' \partial_x\phi_1(x) \partial_{x'}\phi_1(x') \ln \big| e^{i\frac{2\pi}{L}x}-e^{i\frac{2\pi}{L}x'} \big| .
\end{split}
\end{equation}
This time, we have a logarithmic interaction of the field $\partial_x \phi_1$, which is related to the density fluctuation.

\subsubsection{Case of $m_\pm=0$} \label{sec:nchTLL_gapless}

Finally, we consider the case in which both the symmetric and antisymmetric channels are gapless. 
Furukawa and Kim\cite{FurukawaKim:prb11prb13} have studied the R\'enyi entanglement entropy $S_n$ (with $n=1,2,3,\dots$) of this system 
by using the path integral formalism and boundary CFT. 
They have found that R\'enyi entropy obeys a linear function of the chain length $L$
followed by a universal subleading constant determined by $K_+/K_-$. 
Chen and Fradkin\cite{Fradkin_Ladder} have recently studied the ES of this system 
and found that it shows a flat dispersion. 
Here we reproduce these results for a consistency check with these references and for completeness of our discussions. 
We also present the entanglement Hamiltonian written in terms of the fields $\phi_1$ and $\theta_1$. 

For the purpose of comparing with the above references, we define
\begin{equation}
 \kappa = \frac{K_--K_+}{K_-+K_+}, ~~
 z_+ = \frac{-1+\sqrt{1-\kappa^2}}{\kappa}. 
\end{equation}
Setting $m_\pm=0$, we find
\begin{equation}
\begin{split}
 &\cosh(2\theta_{k,\pm}) = \frac12 \left( \frac{K_\pm}{K} + \frac{K}{K_\pm} \right), \\
 &\sinh(2\theta_{k,\pm}) = \frac12 \left( \frac{K_\pm}{K} - \frac{K}{K_\pm} \right).     
\end{split}
\end{equation}
Since these are independent of $k$, 
so are $u_k$, $v_k$, $w_k$, and $f_k$. 
Dropping the subscript $k$ in these, we obtain
\begin{subequations}
\begin{align}
 &u+v=\frac{K w}{2f} (K_+^{-1} + K_-^{-1}), \label{eq:u+v}\\
 &u-v=\frac{w}{2Kf} (K_++K_-), \label{eq:u-v}\\
 &f=\frac1{\sqrt{1-\kappa^2}} = \frac{1-z_+^2}{1+z_+^2},\\
 &w = \ln \frac{1+\sqrt{1-\kappa^2}}{1-\sqrt{1-\kappa^2}} = -\ln z_+^2. \label{eq:flat_ncTLL} 
\end{align}
\end{subequations}
Remarkably, the dispersion $w_k$ is completely flat. 
Since $N_{1,2}=M_{1,2}=0$ in the ground state of the coupled system in this case, 
the zero-mode part \eqref{eq:ve_Ke} disappear in the entanglement Hamiltonian. 
The entanglement Hamiltonian is therefore obtained as
\begin{equation}\label{eq:flatHam_ncTLL}
 H_\ent = w \sum_{k\ne 0} \left( b_{k,1}^\dagger b_{k,1} + \frac12 \right). 
\end{equation}
Using Eqs.~\eqref{eq:u+v} and \eqref{eq:u-v}, this can also be written as
\begin{equation}
\begin{split}
 H_\ent 
  =- &\frac{w(K_+^{-1}+K_-^{-1})}{4\pi f} \\
 &\times\int\!\! \int dx dx' \partial_x\phi_1(x) \partial_{x'}\phi_1(x') \ln \big| e^{i\frac{2\pi}{L}x}-e^{i\frac{2\pi}{L}x'} \big| \\
    - &\frac{w(K_+ +K_-)}{4\pi f} \\
 &\times\int\!\! \int dx dx' \partial_x\theta_1(x) \partial_{x'}\theta_1(x') \ln \big| e^{i\frac{2\pi}{L}x}-e^{i\frac{2\pi}{L}x'} \big|. \\ 
\end{split}
\end{equation}
This time, long-range logarithmic interactions appear both in $\partial_x\phi_1$ and $\partial_x\theta_1$. 

To obtain the Renyi entanglement entropy $S_n$ with $n=2,3,\dots$, 
we first calculate the partition function constructed from $H_\ent$ at the fictitious inverse temperature $\beta$: 
\begin{equation}
 Z_\ent (\beta) = \Tr~ e^{-\beta H_\ent} = \prod_{k\ne 0} \left( 2\sinh\frac{\beta w}{2} \right)^{-1}. 
\end{equation}
We have obtained an infinite product of a constant, which needs to be regularized. 
We introduce a short-distance cutoff $\alpha$, which is of the order of the lattice spacing. 
Then, $k$ runs over $L/\alpha-1$ modes in the above product, 
where the subtraction of $1$ comes from the exclusion of $k=0$. 
We therefore obtain 
\begin{equation}
 Z_\ent (\beta) = \left( 2\sinh\frac{\beta w}{2} \right)^{-\left(\frac{L}{\alpha}-1\right)}.
\end{equation}
The Renyi entanglement entropy is calculated as
\begin{equation}\label{eq:Sn_gapless}
\begin{split}
 S_n 
 &= \frac{-1}{n-1} \ln \frac{Z_\ent (n)}{[Z_\ent (1)]^n} \\
 &= \left( \frac{L}{\alpha} -1 \right) \frac{1}{n-1} \ln \frac{2\sinh\frac{nw}{2}}{\left(2\sinh\frac{w}{2} \right)^n}. 
\end{split}
\end{equation}
As expected, the Renyi entropy $S_n$ is a linear function of the chain length $L$, followed by a subleading constant term. 
The constant term, which we denote by $\gamma_n$, can be rewritten as
\begin{equation}
 \gamma_n 
 = \frac{-1}{n-1} \ln \frac{2\sinh\frac{nw}{2}}{\left(2\sinh\frac{w}{2} \right)^n}
 = \frac{-1}{n-1} \ln \frac{1-(z_+^2)^n}{(1-z_+^2)^n}, 
\end{equation}
which is consistent with Ref.~\onlinecite{FurukawaKim:prb11prb13}. 
The von Neumann entanglement entropy $S_1$ is obtained by taking the limit $n\to 1$ in this expression.\cite{FurukawaKim:prb11prb13}

\section{Numerical analysis} \label{sec:numerics}
%
\begin{figure*}
\begin{center}
\includegraphics[width=0.9\textwidth]{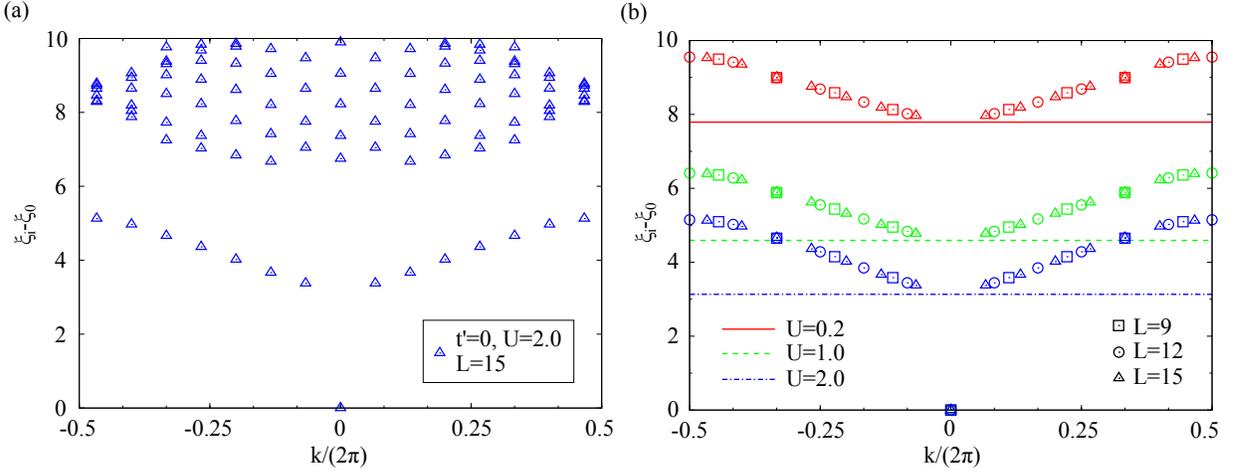}
\end{center}
\caption{(Color online) 
Entanglement excitation spectra $\{\xi_i-\xi_0\}$ plotted against the subsystem momentum $k$ 
in the ladder model \eqref{eq:Boson_Ladder} with $t=1$, $V=-1$, $t'=0$, and varying $U$ at the $1/3$-filling. 
(a) The spectrum for $U=2$ and $L=15$. 
(b) The lowest branch of the spectrum for $U=0.2$ (red), $1.0$ (green), and $2.0$ (blue) from top to bottom. 
The data for different system sizes $L=9$ (square), $12$ (circle), and $15$ (triangle) are plotted together. 
Horizontal lines indicate the energies of the flat dispersion, $w$ in Eq.~(\ref{eq:flat_ncTLL}), predicted by field theory in Sec.~\ref{sec:nchTLL_gapless}. 
They are evaluated to be $w=7.79$, $4.59$, and $3.13$ for $U=0.2$, $1.0$, and $2.0$, respectively, 
using the TLL parameters $K_\pm$ determined in Ref.~\onlinecite{FurukawaKim:prb11prb13}. 
}
\label{fig:ES_2TLL}
\end{figure*}


In the previous section, we have seen through field-theoretical calculations that 
the ES in two coupled non-chiral TLLs displays various low-energy features depending crucially on the interchain couplings. 
In this section, we test these predictions in a numerical diagonalization analysis of a ladder model. 
There have been 
numerical studies on gapped phases of spin ladders,\cite{Poilblanc:prl10, Lauchli:prb12}
for which 
Ref.~\onlinecite{Qi:prl12} and our result in Sec.~\ref{sec:nchTLL_cs_gapped} provides qualitative explanations. 
We therefore focus on gapless phases in the following analysis. 
To this end, we consider a model of hard-core bosons on a ladder
\begin{equation}\label{eq:Boson_Ladder}
\begin{split}
 H =
 &\sum_{\nu=1,2} \sum_{j=1}^L \Big[ -t \left( b_{j,\nu}^\dagger b_{j+1,\nu} + {\rm H.c.} \right) \\
 &~~~~~~~~~~~~~~+V n_{j,\nu} n_{j+1,\nu} -\mu n_{j,\nu} \Big] \\
 &+ \sum_{j=1}^L \left[ -t' \left( b_{j,1}^\dagger b_{j,2} + {\rm H.c.} \right) +U n_{j,1} n_{j,2} \right], \\
\end{split}
\end{equation}
where $b_{j,\nu}$ is a bosonic annihilation operator at the site $j$ 
on the $\nu$th leg,\footnote{We note that $b_{j,\nu}$ introduced here is not related to $b_{k,1}$ used in the previous section.} 
$n_{j,\nu} = b_{j,\nu}^\dagger b_{j,\nu}$ is a number operator defined from it, 
and $L$ is the length of the leg in unit of the lattice spacing. 
Here, $t$ ($t'$) and $V$ ($U$) are the hopping amplitude and the interaction, respectively, between nearest-neighbor sites along the leg (the rung). 
We impose a hard-core constraint $b_{j,\nu}^2=(b_{j,\nu}^\dagger)^2=0$ and periodic boundary conditions ($b_{L+1,\nu}=b_{1,\nu}$). 
This model is thus equivalent to an XXZ model on a ladder by replacing hard-core bosons by spin-$\frac12$ degrees of freedom. The ground-state phase diagram of this model for $t,U>0$ and $-2<V/t\le 0$ has been studied by Takayoshi {\it et al.}\cite{Takayoshi:pra10} The model with $t'=0$ has been used for the numerical test of Eq.~\eqref{eq:Sn_gapless} in Ref.~\onlinecite{FurukawaKim:prb11prb13}. 
As explained in Ref.~\onlinecite{Takayoshi:pra10,FurukawaKim:prb11prb13}, 
for $t'=0$, this model is also equivalent to a fermionic Hubbard chain (with a spin-dependent nearest-neighbor interaction $V$) 
via a Jordan-Wigner transformation. 
Hereafter, we set $t=1$ and $V=-1$, and fix the average density to the $1/3$-filling, $\rho_0 = \left< n_{j,\nu} \right> = 1/3$. 


As discussed in Ref.~\onlinecite{Takayoshi:pra10}, the low-energy effective Hamiltonian of Eq.~(\ref{eq:Boson_Ladder}) is given by
\begin{equation}
\begin{split}
H =& \int dx  \frac{v_+}{2} \left[ K_+ \left( \partial_x \theta_+ \right)^2 + \frac{1}{K_+} \left( \partial_x \phi_+ \right)^2 \right]  \\
   & + \int dx \left\{ \frac{v_-}{2} \left[ K_- \left( \partial_x \theta_- \right)^2 + \frac{1}{K_-} \left( \partial_x \phi_- \right)^2 \right] \right. \\ 
   & ~~~~\left. -C t' \cos \left( \sqrt{2\pi} \theta_- \right) \right\} +\cdots, 
\end{split}
\end{equation}
where $C$ is a constant which depends on $V/t$ and $U/t$.  
We note that the term $U \partial_x \phi_+$ in Ref.~\onlinecite{Takayoshi:pra10} was absorbed into the definition of $\phi_+$. 
This effective Hamiltonian corresponds to the Hamiltonian of Sec.~\ref{sec:nchTLL} 
[Eqs.~\eqref{eq:Hp+Hm} and \eqref{eq:sine-Gordon_pm} in the normalization given by Eq. \eqref{eq:r_rt}]
with $g_+=0$ and $g_-=-Ct'$. 
Therefore, we obtain the following two cases. 
(i) For $t'=0$, the system is described by a two-component TLLs with $m_\pm=0$ (case of Sec.~\ref{sec:nchTLL_gapless}),  
as far as $U$ is below a certain critical value $U_c>0$ at which $K_-$ diverges and a phase transition occurs.\cite{Takayoshi:pra10} 
(ii) For $t'\ne 0$, the cosine term with the scaling dimension $1/(2 K_-)$ ($<1/2$ always in the present analysis) 
opens a finite gap $m_->0$ in the antisymmetric channel (case of Sec.~\ref{sec:nchTLL_c_gapped}). 
Below we introduce the entanglement cut between the two legs, 
obtain the reduced density matrix $\rho_A$ for the first leg,  
and plot the ES against the subsystem momentum $k\in\frac{2\pi}{L}\mathbb{Z}$. 

\begin{figure*}
\begin{center}
\includegraphics[width=0.9\textwidth]{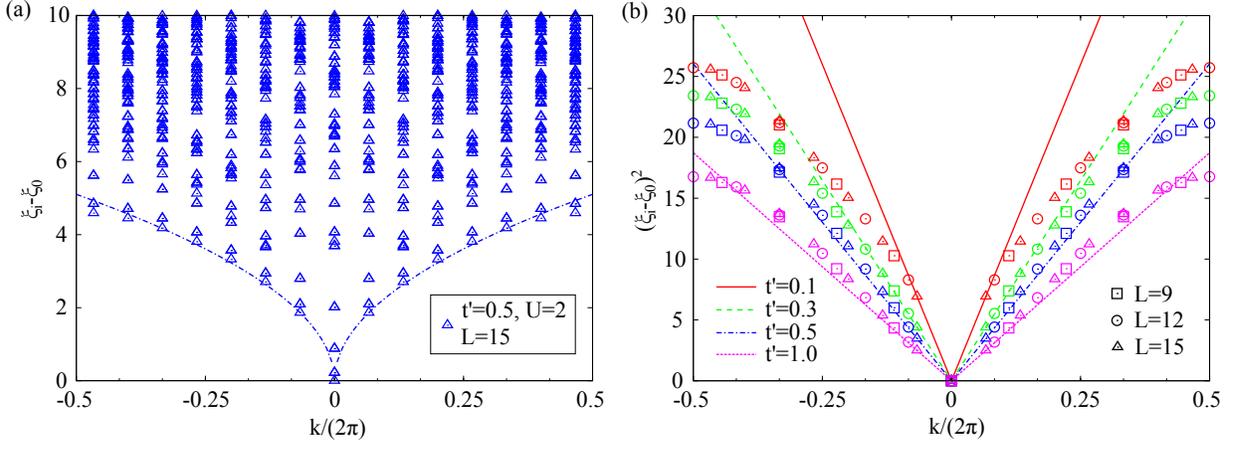}
\end{center}
\caption{(Color online) 
Entanglement excitation spectra $\{\xi_i-\xi_0\}$ in the ladder model \eqref{eq:Boson_Ladder} with $t=1$, $V=-1$, $U=2$ , and varying $t'$ at the $1/3$-filling. 
(a) The spectrum for $t'=0.5$, and $L=15$. 
(b) Squared excitation energies $\{ (\xi_i-\xi_0)^2 \}$ of the lowest branch 
for $t'=0.1$ (red), $0.3$ (green), $0.5$ (blue), and $1.0$ (magenta) from top to bottom. 
The data for different system sizes $L=9$ (square), $12$ (circle), and $15$ (triangle) are plotted together. 
The linear lines are drawn by connecting the origin with the data points at the smallest $|k|\ne 0$. The dashed line in (a) is obtained by simply taking the square root of the linear line in (b).
}
\label{fig:ES_1TLL}
\end{figure*}

\begin{figure}
\begin{center}
\includegraphics[width=0.45\textwidth]{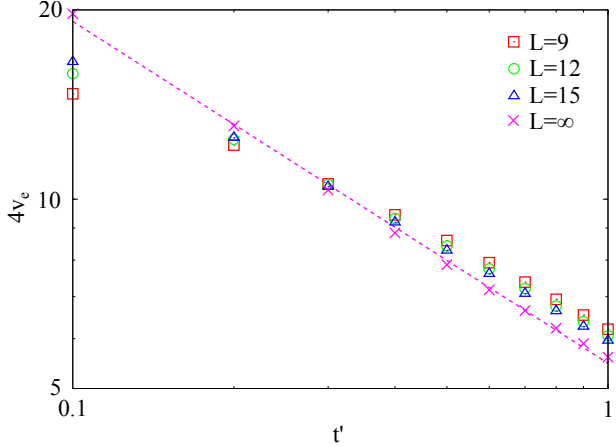}
\end{center}
\caption{(Color online) 
The slope $4v_\ent$ of the linear relation $(\xi_i-\xi_0)^2 \propto |k|$ found in Fig.~\ref{fig:ES_1TLL}(b), 
plotted against the interchain tunneling $t'$. 
Logarithmic scales are taken for both axes. 
We fix $U=2$ and plot the data for the system sizes $L=9$ (square), $12$ (circle), and $15$ (triangle). 
For each $L$, we determine the slope $4v_\ent$ by using the data at $k=2\pi/L$ as shown by lines in Fig.~\ref{fig:ES_1TLL}(b). 
We extrapolate the data of $v_\ent$ to the limit $L\to\infty$ by fitting with a linear function in $1/L$; 
the obtained values are indicated by cross symbols. 
We fit the $L=\infty$ data by a power function (as shown by a dashed line), 
obtaining $4v_\ent=5.48(t')^{-0.544}$. 
The exponent $0.544$ of this function is close to the expected value $1/(2-1/(2K_-))=0.558$. 
}
\label{fig:Slope}
\end{figure}


We first consider the case of $t'=0$, 
for which the result of Sec.~\ref{sec:nchTLL_gapless} predicts an entanglement Hamiltonian with a flat dispersion as in Eq.~(\ref{eq:flatHam_ncTLL}). 
The R\'enyi entanglement entropy $S_n$ in this case has been numerically calculated in Ref.~\onlinecite{FurukawaKim:prb11prb13},   
and the obtained numerical data have been found to agree well with the analytical prediction in Eq.~\eqref{eq:Sn_gapless}. 
Here we analyze the ES $\{ \xi_i \}$, which is obtained from the eigenvalues $\{\exp(-\xi_i)\}$ of the reduced density matrix $\rho_A$. 
Subtracting the lowest entanglement energy $\xi_0$, 
we plot the entanglement excitation spectrum 
$\{\xi_i-\xi_0\}$ for $U=2$ and $L=15$ in Fig.~\ref{fig:ES_2TLL} (a).  
Figure~\ref{fig:ES_2TLL}(b) presents the lowest branch of the spectrum for several values of $U$; 
here the data for three different system sizes $L=9$, $12$ and $15$ are displayed together. 
The lowest branch is expected to correspond to the single-boson excitations in the entanglement Hamiltonian (\ref{eq:flatHam_ncTLL}); 
higher-energy states are expected to come from multi-boson excitations. 
For all the cases investigated, we find that there is a clear excitation gap above $\xi_0$ 
and that the lowest branch has a relatively narrow bandwidth. 
Using the TLL parameters $K_\pm$ obtained numerically in Ref.~\onlinecite{FurukawaKim:prb11prb13}, 
we evaluate the expected energy of the flat dispersion, $w$ in Eq.~(\ref{eq:flat_ncTLL}), 
and plot it by horizontal lines in Fig.~\ref{fig:ES_2TLL} (b). 
We note that our theoretical prediction in Sec.~\ref{sec:nchTLL} was based on the TLL theory, 
which addressed properties for small $|k|$ (below a certain ultraviolet cutoff). 
We find that the numerical data agree well with the horizontal lines around $k=0$, 
and deviate gradually from them as $|k|$ increases. 
These deviations are expected to come from certain effects beyond the TLL theory, 
such as the presence of irrelevant operators. 
Good agreement of the data of $S_n$ with Eq.~\eqref{eq:Sn_gapless} found in Ref.~\onlinecite{FurukawaKim:prb11prb13} 
indicates that such deviations from the TLL prediction at high energies change only the boundary-law contribution 
in $S_n$ and keep the universal constant $\gamma_n$ unaltered. 


We next consider the case of $t' \neq 0$, 
for which the result of Sec.~\ref{sec:nchTLL_c_gapped} predicts an entanglement Hamiltonian 
with an anomalous dispersion relation $w_k$ proportional to $\sqrt{k}$ as in Eq.~(\ref{eq:wk_c_gapped}). 
We here fix the interchain interaction $U=2$ and vary the interchain tunneling $t'$ from $0.1$ to $1$. 
Figure~\ref{fig:ES_1TLL} (a) presents the entanglement excitation spectra for $t'=0.5$ and $L=15$. 
We find that the entanglement excitation energies are significantly lowered around $k=0$ compared to Fig.~\ref{fig:ES_2TLL} (a), indicating a gapless ES. 
Extracting the lowest branch from the spectrum, 
we plot squared excitation energies $\{ (\xi_i-\xi_0)^2 \}$ for several values of $t'$ in Fig.~\ref{fig:ES_1TLL} (b). 
The data roughly show a linear relation $(\xi_i-\xi_0)^2\propto |k|$ (particularly for small $|k|$ and large $t'$), 
which is consistent with the expected square root dispersion relation (\ref{eq:wk_c_gapped}). 
Here we attribute the better agreement with the linear relation for larger $t'$ 
to the increasing mass $m_-$ and thus the associated decreasing correlation length as a function of $t'$; 
in order to have a consistency with a field theory, it is crucial to make the system size sufficiently larger than the correlation length. 
Furthermore, we evaluate the slope $4v_\ent$ of the linear relation from the data at $k=2\pi/L$ for each $L$ 
and plot it against the tunneling amplitude $t'$ in Fig.~\ref{fig:Slope}. 
We also plot together the values of $4v_\ent$ extrapolated to the thermodynamic limit by a linear fitting with $1/L$. 
According to Sec.~\ref{sec:nchTLL_c_gapped}, the slope $4v_\ent$ is proportional to the inverse of the mass $m_-$. 
Taking the perturbative RG picture from the $t'=0$ case, 
the mass should behave as $m_- \sim (t')^{1/(2-1/(2K_-))}$, where $K_-$ is evaluated at $t'=0$. 
Using $K_-=2.42$ obtained in Ref.~\onlinecite{FurukawaKim:prb11prb13}, 
we obtain the exponent $1/(2-1/(2K_-))=0.558$. 
In Fig.~\ref{fig:Slope}, the data of $v_\ent$ extrapolated to $L\to\infty$ show the power-law behavior in $t'$, as indicated by a dashed line; 
the obtained slope $0.544$ is close to the above value, 
demonstrating the consistency with the field-theoretical prediction.  

\section{Summary and Outlook} \label{sec:summary}


In this paper, we have investigated the ES between two coupled (chiral or non-chiral) TLLs on parallel periodic chains. 
By expanding the interchain interactions to quadratic order in bosonic fields, 
we are able to calculate the ES for both gapped and gapless systems using only methods for free theories. 
For two coupled chiral TLLs in Sec.~\ref{sec:chTLL}, 
we have shown that the entanglement Hamiltonian is precisely proportional to the single-chain TLL Hamiltonian at low energies, 
in both the zero mode and the oscillator modes. 
This provides a simple proof for the correspondence between the ES and the chiral edge-mode spectrum in quantum Hall systems consistent with previous numerical and analytical studies. 
Two coupled non-chiral TLLs in Sec.~\ref{sec:nchTLL}, which are relevant to spin ladders and Hubbard chains,  
can be in general reorganized into symmetric and antisymmetric channels which can each be either gapped or gapless. 
When both the two channels are gapped, we have shown that the entanglement Hamiltonian $H_\ent$ 
has linearly dispersing bosonic modes at low energies, 
which resembles a single-chain TLL $H_1$. 
However, $H_\ent$ is characterized by a renormalized TLL parameter $K_\ent$, 
which is different from the original single-chain one $K$ 
unless some special symmetry [e.g., SU$(2)$ symmetry] enforces $K_\ent=K$. 
When one of the two channels is gapless and the other is gapped, 
the entanglement spectrum shows a  dispersion relation proportional to $\sqrt{k}$, where $k$ is the subsystem momentum. 
We have numerically demonstrated the emergence of this interesting dispersion relation in a model of hard-core bosons on a ladder in Sec.~\ref{sec:numerics}. 


In the studies of the ES of gapless spin ladders, two unusual dispersion relations have been found, 
a flat dispersion obtained by Chen and Fradkin\cite{Fradkin_Ladder} (see also Sec.~\ref{sec:nchTLL_gapless}) 
and $\sqrt{k}$ dispersions obtained in Sec.~\ref{sec:nchTLL_c_gapped} and Sec.~\ref{sec:nchTLL_s_gapped}. 
We have discussed that these are related to certain long-range interactions in the entanglement Hamiltonian. 
We note that in a study of critical Rokhsar-Kivelson-type wave functions in two dimensions, 
the entanglement Hamiltonian was found to be given by a 1D lattice gas having a similar long-range interaction.\cite{Stephan:prb09} 
The presence of critical correlations thus appears to be a key ingredient for obtaining long-range interactions in the entanglement Hamiltonian. 
We also mention that it would be interesting to study coupled 1D critical systems 
with non-integer values of the central charge, such as $c=\frac{1}{2}$, 
and see how the dispersion relation of the ES depends on the central charge of the total system.

The approach presented in this paper has a remarkable versatility in treating coupled TLLs with a variety of interactions. 
It will be interesting to apply our method to an array of coupled TLLs 
to address the entanglement properties of higher-dimensional systems from a 1D point of view 
(see Refs.~\onlinecite{PhysRevB.87.241103} and \onlinecite{2013arXiv1309.2255P} for related approaches with other methods). 
It would also be interesting to extend our work to the ES of higher-spin ladders 
as were studied via perturbation theory by Schliemann and L\"auchli.\cite{2012JSMTE..11..021S} 
What makes this situation very interesting is that 
the energy spectrum of a single chain is gapped for certain spin quantum numbers. 
It would also be worthwhile to extend our approach to treat the ES of the Kugel-Khomskii model 
and see if it could explain the interesting features of the ES between spin and orbit degrees of freedom of the model, 
such as the entanglement gap seen at the $SU(4)$ point.\cite{Lundgren}

\acknowledgments

We thank V. Chua, D. Lorshbough, P. Laurell, G. Fiete, and X.L. Qi for useful discussions. RL was supported by NSF Graduate Research Fellowship award number 2012115499. 
RL also thanks the hospitality of the University of Tokyo
where part of this work was completed under
NSF EAPSI award number OISE-1309560
and Japan Society for the Promotion of Science (JSPS)
Summer Program 2013.
YF, SF, and MO were supported by Leading Graduate Course for Frontiers of Mathematical Science and Physics, and KAKENHI Grant Nos. 25800225 and 25400392
from JSPS, respectively.

\bibliography{EScTLL.bib}

\end{document}